\documentclass[11pt]{article}
\usepackage{epsfig}
\author{O. \"Ogetbil}
\title{A General Study of Ground States of $\mathcal{N}=2$ Supergravity Theories with Symmetric Scalar Manifolds in 5 Dimensions}
\usepackage{amssymb}
\usepackage{amsmath}
\usepackage{verbatim}
\numberwithin{equation}{section}
\begin{document}
 
\vspace*{1.8cm}
\begin{center}
{\bf{\large A General Study of Ground States in Gauged $\mathcal{N}=2$ Supergravity Theories with Symmetric Scalar Manifolds in 5 Dimensions}}  \\

\vspace{1cm}
\begin{large}

 O. \"Ogetbil\footnote{oogetbil@phys.psu.edu}
\end{large}
\\
\vspace{.35cm}

\vspace{.3cm}  \emph{Physics Department \\
Pennsylvania State University\\
University Park, PA 16802, USA} \\

\vspace{1cm}
{\bf Abstract}
\end{center}

 After reviewing the existing results we give an extensive analysis of the critical points of the potentials of the gauged $\mathcal{N}=2$ Yang-Mills/Einstein Supergravity theories coupled to tensor- and hyper multiplets. Our analysis includes all the possible gaugings of all $\mathcal{N}=2$ Maxwell-Einstein supergravity theories whose scalar manifolds are symmetric spaces. In general, the scalar potential gets contributions from $R$-symmetry gauging, tensor couplings and hyper-couplings. We show that the coupling of a hypermultiplet into a theory whose potential has a non-zero value at its critical point, and gauging a compact subgroup of the hyperscalar isometry group will only rescale the value of the potential at the critical point by a positive factor, and therefore will not change the nature of an existing critical point. However this is not the case for non-compact $SO(1,1)$ gaugings. An $SO(1,1)$ gauging of the hyper isometry will generally lead to deSitter vacua, which is analogous to the ground states found by simultaneously gauging $SO(1,1)$ symmetry of the real scalar manifold with $U(1)_R$ in earlier literature. $SO(m,1)$ gaugings with $m>1$, which give contributions to the scalar potential only in the Magical Jordan family theories, on the other hand, do not lead to deSitter vacua. Anti-deSitter vacua are generically obtained when the $U(1)_R$ symmetry is gauged. We also show that it is possible to embed certain generic Jordan family theories into the Magical Jordan family preserving the nature of the ground states. However the Magical Jordan family theories have additional ground states which are not found in the generic Jordan family theories.

\thispagestyle{empty}
\newpage
\tableofcontents\setcounter{page}{1}
\newpage
\section{Introduction}
Higher dimensional gauged supergravity theories have been studied extensively in the eighties \cite{Salam:1989fm}. These include the five dimensional gauged supergravity theories \cite{Gunaydin:1984ak},\cite{Gunaydin:1985cu} that received renewed attention more recently due their role within the AdS/CFT correspondences in string theory \cite{Maldacena:1997reandtheothers}, Randall-Sundrum (RS) braneworld scenario \cite{Randall:1999ee} and M-theory compactifications on Calabi-Yau threefolds with fluxes \cite{Cadavid:1995bk}. It is believed that the $5D, \mathcal{N}=8$ gauged supergravity \cite{Gunaydin:1985cu} is a consistent nonlinear truncation of the lowest lying Kaluza-Klein modes of type IIB supergravity on $AdS_5\times S^5$ \cite{Gunaydin:1984fk}. Moreover, certain brane world scenarios based on M-theory compactifications have $5D, \mathcal{N}=2$ gauged supergravity as their effective field theories \cite{Horava:1995qa}. 

Meanwhile, the evidence of a small positive cosmological constant from recent cosmological observations \cite{Perlmutter:1998np} attracted interest in finding stable deSitter ground state solutions in string theory \cite{Kachru:2003aw} and supergravity theories \cite{Hull:2001ii},\cite{Fre:2002pd}. In the context of supersymmetric theories, Anti-deSitter ground states emerge naturally in contrast to deSitter ground states. This is due to the fact that the deSitter superalgebras usually have non-compact $R$-symmetry subalgebras, which leads to existence of ghosts if the supersymmetry is to be fully preserved. Nevertheless exact supersymmetry is not observed in nature and supersymmetry arises as a broken symmetry.

In this paper, we shall focus on $5D, \mathcal{N}=2$ gauged supergravity theories coupled to vector, tensor and hypermultiplets by gauging various symmetries. The analysis is somewhat easier than in $4D$, mainly because in $4D$, the $U$-duality is an on-shell symmetry, whereas in $5D$, it is a symmetry of the Lagrangian. Moreover $5D$ theories have real geometry while the geometry in $4D$ is complex. This limits the possible gaugings and helps us doing an almost complete analysis of critical points in five dimensions. 

Pure $5D, \mathcal{N}=2$ supergravity was constructed in \cite{Cremmer:1980gs}, coupling to vector multiplets was done in \cite{Gunaydin:1983bi},\cite{Gunaydin:1984ak} and tensor fields were added to the theory in \cite{Gunaydin:1999zx}. Coupling of hypers to these theories was done in \cite{Ceresole:2000jd}. Vacua of $U(1)_R$ gauged $5D, \mathcal{N}=2$ Maxwell-Einstein supergravity theories (MESGT's) and Yang-Mills/Einstein supergravity theories (YMESGT's) without hypers and tensors were studied in \cite{Gunaydin:1984ak}. Vacua of the generic Jordan family models, which will be defined below, with Abelian gaugings and tensors have been investigated in \cite{Gunaydin:Vacua}, the full $R$-symmetry group gauging was done in \cite{Gunaydin:2000ph} and some other possible gaugings have been carried out in \cite{smet}. We will study the vacua of these theories that have and have not been covered in the literature so far, generalize the previous results obtained for a fixed number $\tilde{n}$ of vector multiplets to arbitrary $\tilde{n}$ and investigate the ground states when a universal hypermultiplet coupled to these theories.

We adopt the convention introduced in \cite{Gunaydin:1999zx} to classify the gaugings of supergravity theories. The ungauged $5D,  \mathcal{N}=2$ supergravity coupled to vector- and/or hypermultiplets is referred as (ungauged) MESGT. Theories obtained by gauging a $U(1)_R$ subgroup of $SU(2)_R$ by coupling a linear combination of vector fields to the fermions \cite{Gunaydin:1984ak}, which are the only fields that transform nontrivially under $SU(2)_R$, are called gauged Maxwell-Einstein supergravity theories (gauged MESGT). On the other hand, if only a subgroup $K$ of the symmetry group of the action is being gauged, the theory is referred as a YMESGT. Note that the theories which include tensor fields fall into this category. A theory with a gauge group $K\times U(1)_R$ is called gauged Yang-Mills/Einstein supergravity theory (gauged YMESGT).

The organization of this paper is as follows. In section 2, we review the field content of the $5D, \mathcal{N}=2$ supergravity, its possible gaugings and the potential terms arising from these gaugings. Sections 3,4,5 deal with the ground states of the generic Jordan family, Magical Jordan family and generic non-Jordan family theories, respectively, that are subject to such gaugings that give non-trivial potential terms. The critical points, if they exist, of these theories are given and their stability is discussed. The addition of hypers in the theory generally makes the equations for the stability calculations very complicated. Hence, in certain cases we will just give particular numerical examples that show that it is possible to obtain stable vacua when hypers are coupled to the theory. Section 6 collects the summary of all the novel ground states found in this paper, as well as the previously known results. In the first appendix, one can find the bosonic part of the Lagrangian, the elements of the very special geometry and the derivation of the potential terms from more fundamental quantities. Appendix B lists the Killing vectors and their corresponding prepotentials of the hyperscalar manifold isometries that will be used to carry out the hyper-gaugings throughout the paper.
 
\section{The Basics}
\setcounter{equation}{0}
In this section an outline of the theory to start with  and our comventions will be given. The potentials of $\mathcal{N}=2$ supergravity theories coupled to tensor- and/or hypermultiplets will be reviewed.\\[2pt]
The field content of the ungauged (before tensor- or hypermultiplet coupling) $\mathcal{N}=2$ MESGT is
\begin{equation}
\{e^m_\mu,\Psi^i_\mu,A^I_\mu,\lambda^{i\tilde{a}},\varphi^{\tilde{x}}\}
\end{equation}
where
\begin{eqnarray}
\begin{array}{rcl}
i&=&1,2\\
I&=&0,1,...,\tilde{n},\\
\tilde{a}&=&1,2,...,\tilde{n}\\
\tilde{x}&=&1,2,...,\tilde{n}\nonumber
\end{array}
\end{eqnarray}
The 'graviphoton' is combined with the $\tilde{n}$ vector fields of the $\tilde{n}$ vector multiplets into a single ($\tilde{n}$+1)-plet of vector fields $A^I_\mu$ labelled by the index $I$. The indices $\tilde{a},\tilde{b},...$ and $\tilde{x},\tilde{y},...$ are the flat and the curved indices, respectively, of the $\tilde{n}$-dimensional target manifold $\mathcal{M}_{VS}$ of the real scalar fields, which we will define below.\\[2pt]
The bosonic part of the Lagrangian is given in the Appendix A. The global symmetries of these theories are of the form $SU(2)_R \times G$, where $SU(2)_R$ is the $R$-symmetry group of the $\mathcal{N}=2$ Poincare superalgebra and $G$ is the subgroup of the group of isometries of the scalar manifold that extends to the symmetries of the full action. Gauging a subgroup $K$ of $G$ requires dualization of some of the vector fields to self-dual tensor fields if they are transforming in a non-trivial representation of $K$. More formally, the field content, when $m$ of the vector fields are dualized to tensor fields, becomes
\begin{equation}
\{e^m_\mu,\Psi^i_\mu,A^I_\mu,B^M_{\mu\nu},\lambda^{i\tilde{a}},\varphi^{\tilde{x}}\}
\end{equation}
where now
\begin{eqnarray}
\begin{array}{rcl}
i&=&1,2\\
I&=&0,1,...,n\\
M&=&1,2,...,2m\\
\tilde{I}&=&0,1,...,\tilde{n}\\
\tilde{a}&=&1,2,...,\tilde{n}\\
\tilde{x}&=&1,2,...,\tilde{n}
\end{array}\nonumber
\end{eqnarray}
with $\tilde{n}=n+2m$. Tensor multiplets come in pairs with four spin-$1/2$ fermions (i.e. two $SU(2)_R$ doublets) and two scalars. Tensor coupling generally introduces a scalar potential of the form \cite{Gunaydin:1999zx}:

\begin{equation}
P^{(T)}=\frac{3\sqrt{6}}{16}h^I\Lambda^{MN}_I h_M h_N.\label{pt}
\end{equation}
Here $\Lambda^{MN}_I$ are the transformation matrices of the tensor fields and $h_{\tilde{I}}, h^{\tilde{I}}$ are elements of the ``very special'' geometry of the scalar manifold $\mathcal{M}_{VS}$ that has the metric $\stackrel{o}{a}_{\tilde{I}\tilde{J}}$ which is used to raise and lower the indices $\tilde{I},\tilde{J}...$.  

When the full $R$-symmetry group $SU(2)_R$ is being gauged the potential gets the contribution
\begin{equation}
 P^{(R)}=-4 C^{AB\tilde{K}}\delta_{AB} h_{\tilde{K}},\label{su2r}
\end{equation}
where $A,B$ are adjoint indices of $SU(2)$. If instead, the $U(1)_R$ subgroup is being gauged, the contribution to the potential becomes
\begin{equation}
P^{(R)}=-4C^{IJ\tilde{K}}V_I V_J h_{\tilde{K}}.\label{u1r}
\end{equation}
The expressions that lead to the derivation of the above potential terms can be found in the Appendix A.

We will look at the cases, where the scalar manifold $\mathcal{M}_{VS}$ is a symmetric space. Such spaces are further divided in two categories, depending whether they are associated with a Jordan algebra or not. The spaces that are associated with Jordan algebras are of the form $\mathcal{M}_{VS}=\frac{Str_0 (J)}{Aut(J)}$, where $Str_0 (J)$ and $Aut(J)$ are the reduced structure group and the automorphism group, respectively, of a real, unital Jordan algebra $J$, of degree three \cite{Gunaydin:1983bi}, \cite{Gunaydin:1986fg}. More specifically,
\begin{itemize}
\item Generic Jordan Family:
\begin{equation}\begin{array}{rlr}
J=\mathbb{R}\oplus \Sigma_{\tilde{n}}:&\displaystyle \mathcal{M}_{VS}=\frac{SO(\tilde{n}-1,1)\times SO(1,1)}{SO(\tilde{n}-1)},&\tilde{n}\geq 1.\nonumber
\end{array}\end{equation}
\item Magical Jordan Family:
\begin{equation}\begin{array}{rll}
J_3^{\mathbb{R}}:\quad&\displaystyle \mathcal{M}_{VS}=\frac{SL(3,\mathbb{R})}{SO(3)},&\quad\tilde{n}=5,\\[9pt]
J_3^{\mathbb{C}}:\quad&\displaystyle \mathcal{M}_{VS}=\frac{SL(3,\mathbb{C})}{SU(3)},&\quad\tilde{n}=8,\\[9pt]
J_3^{\mathbb{H}}:\quad&\displaystyle \mathcal{M}_{VS}=\frac{SU^{*}(6)}{Usp(6)},&\quad\tilde{n}=14,\\[9pt]
J_3^{\mathbb{O}}:\quad&\displaystyle \mathcal{M}_{VS}=\frac{E_{6(-26)}}{F_4},&\quad\tilde{n}=26.\nonumber
\end{array}\end{equation}
\item Generic non-Jordan Family:
\begin{equation}
\mathcal{M}_{VS}=\frac{SO(1,\tilde{n})}{SO(\tilde{n})},\quad \tilde{n}\geq 1.\nonumber
\end{equation}
\end{itemize}

In addition to the supergravity multiplet, $n$ vector multiplets and $m$ tensor multiplets one can couple hypermultiplets into the theory. A universal hypermultiplet
\begin{equation}
\{\zeta^a,q^X\}
\end{equation}
contains a spin-1/2 fermion doublet $A=1,2$ and four real scalars $X=1,...,4$. The total manifold of the scalars $\phi=(\varphi,q)$ then becomes
\begin{equation}
\mathcal{M}_{scalar} = \mathcal{M}_{VS} \otimes \mathcal{M}_{Q}\nonumber
\end{equation}
with $dim_{\mathbb{R}}\mathcal{M}_{VS}=n+m$ and $dim_{\mathbb{Q}}\mathcal{M}_{Q}=1$. The quaternionic hyperscalar manifold $\mathcal{M}_{Q}$ of the scalars of a single hypermultiplet has the isometry group $SU(2,1)$. Gauging a subgroup of this group introduces an extra term in the scalar potential \cite{Ceresole:2000jd}
\begin{equation}
P^{(H)}=2\mathcal{N}_{iA}\mathcal{N}^{iA}
\end{equation}
where $\mathcal{N}^{iA}=\frac{\sqrt{6}}{4} h^I K^X_I f^{iA}_X$ with $f^{iA}_X$ being the quaternionic vielbeins, $f^{iA}_X f_{YiA}=g_{XY}$, and $g_{XY}$ is the metric of the quaternionic-Kahler hypermultiplet scalar manifold \cite{Ceresole:2001wi}
\begin{equation}
ds^2=\frac{dV^2}{2V^2}+\frac{1}{2V^2} (d\sigma +2\theta d\tau - 2 \tau d\theta)^2 + \frac{2}{V} (d\tau^2+d\theta^2),
\end{equation}
and $K^X_I$ being the Killing vectors given in the Appendix B together with their corresponding prepotentials. The determinant of the metric is $1/V^6$ and it is positive definite and well behaved everywhere except $V=0$. But since in the Calabi-Yau derivation V corresponds to the volume of the Calabi-Yau manifold\cite{Lukas:1998tt}, we restrict ourselves to the positive branch $V>0$.

When the $R$-symmetry is gauged in a theory that contains hypers, the potential $P^{(R)}$ gets some modification due to the fact that the fermions in the hypermultiplet are doublets under the $R$-symmetry group $SU(2)_R$. It becomes
\begin{equation}
P^{(R)}=-4 C^{IJ\tilde{K}} \vec{P}_I\cdot\vec{P}_J h_{\tilde{K}}\label{pr_h}
\end{equation}
where $\vec{P}_I$ are the prepotentials corresponding to the Killing vectors $K^X_I$.

The total scalar potential, which includes terms from tensor coupling, $R$-symmetry gauging and hyper coupling, is given by
\begin{equation}
\begin{array}{rcl}
e^{-1}\mathcal{L}_{pot}&=&- g^2 P^{(T)} - g_R^2 P^{(R)} - g_H^2 P^{(H)}\\
&\equiv&- g^2 P^{(5)}_{TOT}\\[5pt]
&=&- g^2 (P^{(T)} + \lambda P^{(R)} + \kappa P^{(H)}).\label{totalpot}
\end{array}
\end{equation}
where $\lambda=g_R^2/g^2$, $\kappa=g_H^2/g^2$; $g_R$, $g_H$ and $g$ are coupling constants, which need not to be all independent.
\paragraph{Supersymmetry of the solutions: }
Demanding supersymmetric variations of the fermions vanish at the critical points of the theory, the conditions that need to be satisfied are found as \cite{Gunaydin:Vacua},\cite{Ceresole:2001wi}
\begin{equation}
 \langle W^{\tilde{a}}\rangle = \langle P^{\tilde{a}}\rangle = \langle \mathcal{N}_{iA}\rangle=0
\end{equation}
where $W^{\tilde{a}}$ and $P^{\tilde{a}}$ are defined in (\ref{wapadef}). Any ground state that does not satisfy all of these conditions are not supersymmetric. One can see that any supersymmetric solution must be of the form
\begin{equation}
 P^{(5)}_{TOT}\arrowvert_{\phi^C} = -4 \lambda \text{ }\vec{P} \cdot \vec{P}(\phi^C)
\end{equation}
which is negative semi-definite. Hence we know from beginning that any deSitter type ground state of the theories we will consider will have broken supersymmetry. The parametrization of the Killing vectors of the hyperscalar manifold, which is outlined in the Appendix B, yields $K_I^X\arrowvert_{q^C}\neq 0$, for noncompact generators. Here, the point $q^C=\{V=1,\sigma=\theta=\tau=0\}$ is the base point of the hyperscalar manifold, i.e. the compact Killing vectors of the hyper-isometry generate the isotropy group of this point. This point will be used as the hyper-coordinate candidate of the critical points. As a consequenece $\langle \mathcal{N}_{iA}\rangle\neq 0$; and hence theories including noncompact hyper-gauging will not have supersymmetric critical points either.

\section{Generic Jordan Family}

The theory being considered is $\mathcal{N}=2$ supergravity coupled to $\tilde{n}$ Abelian vector multiplets and with real scalar manifold $\mathcal{M}_{VS}=SO(\tilde{n}-1,1)\times SO(1,1)/SO(\tilde{n}-1),\tilde{n}\ge 1$. The cubic polynomial can be written in the form \cite{Gunaydin:Vacua}
\begin{equation}
N(h)=\frac{3\sqrt{3}}{2}h^0[(h^1)^2 - (h^2)^2 - ... - (h^{\tilde{n}})^2]
\end{equation}
The non-zero $C_{\tilde{I}\tilde{J}\tilde{K}}$'s are
\begin{equation}
C_{011}=\frac{\sqrt{3}}{2},\qquad C_{022}=C_{033}=...=C_{0\tilde{n}\tilde{n}}=-\frac{\sqrt{3}}{2}\nonumber
\end{equation}
and their permutations. The constraint $N=1$ can be solved by
\begin{equation}
  h^0=\frac{1}{\sqrt{3}||\varphi||^2},\qquad h^{a}=\sqrt{\frac{2}{3}}\varphi^a\nonumber
\end{equation}
with $a,b=1,2,..,\tilde{n}$ and $||\varphi||^2 =\varphi^a\eta_{ab}\varphi^{b}$, where $\eta_{xy}=(+--...-)$. The scalar field metric metrics $g_{\tilde{x}\tilde{y}}$ and vector field metric $\stackrel{o}{a}_{\tilde{I}\tilde{J}}$ that appear in the kinetic terms in the Lagrangian are positive definite in the region $||\varphi||^2 >0$. In order to have theories that have a physical meaning, our investigation is restircted to this region. As a consequence one must have $\varphi^1\neq 0$.\\[3pt]

The isometry group of the real scalar manifold $\mathcal{M}_{VS}$ is $G = SO(\tilde{n}-1,1)\times SO(1,1)$. This is the symmetry group of the full action modulo the isometry group of the hyperscalar manifold. The gauging of an $SO(1,1)$ or an $SO(2)$ subgroup of $SO(\tilde{n}-1,1)$ will lead to dualisation of vectors to tensor fields and this gives a scalar potential term. In the generic Jordan family there are no vector fields that are nontrivially charged when the gauge group is non-Abelian, and hence gauging a non-Abelian subgroup of $G$ will not give a scalar potential term. It is also possible to gauge the $R$-symmetry group $SU(2)_R$ or its subgroup $U(1)_R$; or one can introduce a hypermultiplet in the theory and gauge its symmetries to get additional scalar terms in the potential. We will look at each case in turn.
\subsection{Maxwell-Einstein supergravity}
\subsubsection{No $R$-symmetry gauging}
\paragraph{Without hypermultiplets:}
There is no scalar potential and the vacuum is Minkowskian.
\paragraph{With a universal hypermultiplet:}
One can gauge $U(1) \subset SU(2) \times U(1)$ or a non-compact subgroup $SO(1,1)$. To gauge the $U(1)$ symmetry one has to take the Killing vector $\vec{K}$ as a linear combination of $\vec{T}_1, \vec{T}_2, \vec{T}_3$ and $\vec{T}_8$ of (\ref{Tvex}) whereas one has to take a linear combination of $\vec{T}_4,\vec{T}_5,\vec{T}_6$ and $\vec{T}_7$ if he is to gauge the $SO(1,1)$. We take a linear combination of the vector fields ($V_I A_\mu^I$) from the vector multiplet as our gauge field. For $U(1)$ gauging, the corresponding Killing vector and hence the term $P^{(H)}=2\mathcal{N}_{iA}\mathcal{N}^{iA}$ vanishes at its critical point ($V=1, \sigma=\theta=\tau=0$). An important consequence of this is, for the generic family, $U(1)$ gauging of the hyper isometry will not change the sign of the critical points of the theory. Simultaneous gauging of $U(1)\subset SU(2)$ with $U(1)_R$ will only rescale $P_R$ by a positive factor. Such a scaling can be absorbed by redefining $V_I$'s. But the stability of the vacuum will still need to be checked. We will see an example to this in section \ref{so11u1rhypers}.

The situation is slightly different when a non-compact gauging of hyper isometry is done. A linear combination of all vector fields at hand ($A_\mu[SO(1,1)]=V_I A_\mu^I$) is taken as the gauge field. More precisely,
\begin{equation}
 \mathcal{N}^{iA}=\frac{\sqrt{6}}{4} (V_I h^I) (W^k T_k^X) f_X^{iA}\qquad k=4,...,7;\quad I=0,...,\tilde{n}.
\end{equation}
At the base point $q^C=\{V=1, \sigma=\theta=\tau=0\}$ of hyperscalar manifold, one finds
\begin{equation}
\begin{array}{rcl}
 \partial_{\varphi^1} P^{(5)}_{TOT}\arrowvert_{q^C}&=&\frac{1}{4} \kappa \left(2 (W^4)^2+2 (W^5)^2+(W^6)^2+(W^7)^2\right)\left(\sqrt{2} V_1 - \frac{2 V_0 \varphi^1}{||\varphi||^4}\right)\\
&&\left( \sqrt{2}(V_1 \varphi^1 + ... + V_{\tilde{n}} \varphi^{\tilde{n}}) + \frac{V_0}{||\varphi||^2}\right),\end{array}\nonumber
\end{equation}

\begin{equation}
\begin{array}{rcl}
 \partial_{\varphi^a} P^{(5)}_{TOT}\arrowvert_{q^C}&=&\frac{1}{4} \kappa \left(2 (W^4)^2+2 (W^5)^2+(W^6)^2+(W^7)^2\right)\left(\sqrt{2} V_a + \frac{2 V_0 \varphi^a}{||\varphi||^4}\right)\\
&&\left( \sqrt{2}(V_1 \varphi^1 + ... + V_{\tilde{n}} \varphi^{\tilde{n}}) + \frac{V_0}{||\varphi||^2}\right),\qquad\qquad (a=2,...,\tilde{n}),\\
\partial_{V} P^{(5)}_{TOT}\arrowvert_{q^C}&=&\partial_{\sigma} P^{(5)}_{TOT}\arrowvert_{\varphi^C}=0 .\\
\partial_{\theta} P^{(5)}_{TOT}\arrowvert_{q^C}&=&\frac{1}{4} \kappa W^4 W^6 \left( \sqrt{2}(V_1 \varphi^1 + ... + V_{\tilde{n}} \varphi^{\tilde{n}}) + \frac{V_0}{||\varphi||^2}\right)^2 ,\\
\partial_{\tau} P^{(5)}_{TOT}\arrowvert_{q^C}&=&\frac{1}{4} \kappa W^4 W^7 \left( \sqrt{2}(V_1 \varphi^1 + ... + V_{\tilde{n}} \varphi^{\tilde{n}}) + \frac{V_0}{||\varphi||^2}\right)^2 .
\end{array}\nonumber
\end{equation}
These expressions simultaneously vanish by letting
\begin{equation}
 \frac{V_1}{\varphi^1}=-\frac{V_2}{\varphi^2}=...=-\frac{V_{\tilde{n}}}{\varphi^{\tilde{n}}}=\frac{\sqrt{2} V_0}{||\varphi||^4}\label{justhypers}
\end{equation}
and by setting either $W^6=W^7=0$ or $W^4=0$. In the former case, the potential at the critical point becomes
\begin{equation}
 P^{(5)}_{TOT}\arrowvert_{\phi^C} = \frac{6 \kappa \left((W^4)^2+(W^5)^2 \right) (V_0)^2}{4||\varphi||^4}.
\end{equation}
which is positive definite if not all $W^4, W^5, V_0$ are zero. The condition $||\varphi||^2>0$ together with the equation (\ref{justhypers}) determine the constraint on $V_I$'s as
\begin{equation}
 (V_1)^2-(V_2)^2-...-(V_{\tilde{n}})^2>0.
\end{equation}
Stability of this critical point is checked by calculating the Hessian of the potential at the critical point. Using the $SO(\tilde{n}-1,1)$ symmetry one can rotate the fields such that $\varphi^2=...=\varphi^{\tilde{n}}=0$. In particular for $\tilde{n}=3$, the eigenvalues of the Hessian are found to be
\begin{equation}
 \left(0, \tilde{A}, \tilde{A}, 3\tilde{A}, \tilde{A}\tilde{B}, \tilde{A}\tilde{B}, \frac{3\tilde{A}}{2(\varphi^1)^2}\right)
\end{equation}
where
\begin{equation}
 \begin{array}{rcl}
  \tilde{A}&=&\displaystyle \frac{27 \kappa^2 (V_0)^4 \left( (W^4)^2+(W^5)^2 \right)^2 }{4 (\varphi^1) ^{10}}\\[6pt]
\tilde{B}&=&\frac{3}{4}(\varphi^1)^2\left((W^4)^2+(3 W^5)^2 \right)^2 .
 \end{array}
\end{equation}

These eigenvalues are all non-negative, hence the critical point of the potential corresponds to a stable deSitter vacuum.

The same result can be obtained by letting $W^4=0$ instead of $W^6=W^7=0$.

\subsubsection{$SU(2)_R$ symmetry gauging}
In order to have $SU(2)\sim SO(3)$ to be a subgroup of the isometry $SO(\tilde{n}-1,1)\times SO(1,1)$, one obviously needs $\tilde{n}\geq 4$.
\paragraph{Without hypermultiplets:}
The calculation has been done in \cite{Gunaydin:2000ph} with $A_\mu^2, A_\mu^3, A_\mu^4$ taken as gauge fields and the potential was found to be
\begin{equation}
P^{(5)}_{TOT}=\lambda P^{(R)}=6\lambda ||\varphi||^2.
\end{equation}
This potential does not have any critical points\footnote{One can take $\lambda=0$ but this will make the potential vanish everywhere} in the physically relevant region $||\varphi||^2>0$.
\paragraph{With a universal hypermultiplet:}
The gauging of $SU(2)_R$ must be done simultaneously with the gauging of $SU(2) \subset SU(2,1)$ of the hyperscalar manifold. Hence one has $\lambda=\kappa$ in this case. Without loss of generality, one can choose $A_\mu^2, A_\mu^3$ and $A_\mu^4$ as our gauge fields and identify the Killing vectors as
\begin{equation}
K_2^X = T_1^X \qquad K_3^X = T_2^X \qquad K_4^X = T_3^X\label{su2kt}
\end{equation}
and the prepotentials are taken accordingly. The scalar potential is now
\begin{equation}
P^{(5)}_{TOT}=\lambda (P^{(R)}+P^{(H)})
\end{equation}
with $P^{(R)}$ defined as in (\ref{pr_h}).
The derivative of the total potential with respect to $\varphi^1$ is given by
\begin{equation}\begin{array}{rcl}
\frac{\partial P^{(5)}_{TOT}}{\partial \varphi^1}&=& \lambda \Big\{ \Big(V^4+4 \left(\theta ^2+\tau ^2+11\right) V^3 \\
&&+2 \left(3 \theta ^4+\left(6 \tau ^2+46\right)
   \theta ^2+3 \tau ^4+\sigma ^2+46 \tau ^2+51\right) V^2\\
&&+4 \left(\theta ^2+\tau ^2+11\right) \left(\theta ^4+2
   \left(\tau ^2+1\right) \theta ^2+\sigma ^2+\left(\tau ^2+1\right)^2\right) V\\
&&+\left(\theta ^4+2 \left(\tau
   ^2+1\right) \theta ^2+\sigma ^2+\left(\tau ^2+1\right)^2\right)^2\Big) \varphi^1 \Big\} /(32 V^2)\\
&\stackrel{q^C}{\longrightarrow}& 6\lambda \varphi^1
\end{array}\nonumber\end{equation}
and it cannot be brought to zero in the physically relevant region, unless if $\lambda=0$, but that turns off the potential and leads to Minkowski vacuum, hence the potential has no critical points for this case. However one can gauge an additional $U(1)$ and/or $SO(1,1)$ symmetry of the hyperscalar manifold to have extra contributions to the scalar potential. 
\subparagraph{$SU(2)_R \times U(1)_H$ gauging:}
A similar situation occurs as in the last case. The potential has no critical points.
\subparagraph{$SU(2)_R \times SO(1,1)_H$ gauging:}
We choose the linear combination $V_b A^b_\mu$, $b=0,1,5,$ $6,...,\tilde{n}$ as the $SO(1,1)$ gauge field and the noncompact $T_4^X$ as the Killing vector for this gauging. The potential is given by 
\begin{equation}
P^{(5)}_{TOT}=\lambda (P^{(R)}+2\mathcal{N}_{iA}\mathcal{N}^{iA})
\end{equation}
where now $\mathcal{N}^{iA}=\frac{\sqrt{6}}{4} (h^a K_a^X+ (V_b h^b) T_4^X) f^{iA}_X$ with $a=2,3,4$; $K^X_a$ were defined in (\ref{su2kt}); the coupling constant for the $SO(1,1)$ gauging is absorbed in $V_b$'s.

At the base point of the hyperscalar manifold $q^C=\{V=1, \sigma=\theta=\tau=0\}$ the derivatives of the potential are evaluated as,
\begin{equation}\begin{array}{rcl}
 \partial_{\varphi^1} P^{(5)}_{TOT}\arrowvert_{q^C}&=&\lambda\left( 3 \varphi^1 +\frac{1}{\sqrt{2}}\left( V_1 - \frac{\sqrt{2} V_0 \varphi^1}{||\varphi||^4}\right) \tilde{C}\right),\\
 \partial_{\varphi^a} P^{(5)}_{TOT}\arrowvert_{q^C}&=&\lambda \varphi^a \left(-3+\frac{V_0 \tilde{C}}{||\varphi||^4}\right),\\
 \partial_{\varphi^d} P^{(5)}_{TOT}\arrowvert_{q^C}&=&\lambda\left( -3 \varphi^d +\frac{1}{\sqrt{2}}\left( V_d + \frac{\sqrt{2} V_0 \varphi^d}{||\varphi||^4}\right) \tilde{C}\right),\qquad d=5,6,...,\tilde{n},\\
\partial_{V} P^{(5)}_{TOT}\arrowvert_{q^C}&=&0 \\
\partial_{\sigma} P^{(5)}_{TOT}\arrowvert_{q^C}&=&\frac{\sqrt{2}}{4}\lambda \varphi^4 \tilde{C}\\
\partial_{\theta} P^{(5)}_{TOT}\arrowvert_{q^C}&=& \frac{\sqrt{2}}{2}\lambda \varphi^3 \tilde{C}\\
\partial_{\tau} P^{(5)}_{TOT}\arrowvert_{q^C}&=&\frac{\sqrt{2}}{2}\lambda \varphi^2 \tilde{C}
\end{array}\label{hypersu2so11der}\end{equation}
where
\begin{equation}
 \tilde{C}=\sqrt{2}(V_e \varphi^e)+\frac{V_0}{||\varphi||^2},\qquad e=1,5,6,...,\tilde{n}.\nonumber
\end{equation}
In order to set the last three equations of (\ref{hypersu2so11der}) to zero one might set $\tilde{C}=0$, but applying this to the first equation makes it impossible to vanish, unless $\lambda=0$, but that makes the overall potential zero. Hence we set $\varphi^2_C=\varphi^3_C=\varphi^4_C=0$. Then all left to solve are the first and the third equations. Motivated by (\ref{justhypers}) we set
\begin{equation}
V_1 \varphi^d = - \varphi^1 V_d\qquad \forall d=5,6,...,\tilde{n}.\label{vieqnssu2hyp}
\end{equation}
This reduces the first and third equations of (\ref{hypersu2so11der}) to
\begin{equation}
 \lambda \varphi^e \left( 3+\frac{(-2 V_0 \varphi^1 +\sqrt{2} V_1 ||\varphi||^4) (V_0 \varphi^1 +\sqrt{2} V_1 ||\varphi||^4)}{2(\varphi^1)^2 ||\varphi||^6} \right)=0
\end{equation}
Solving this for $\varphi^1$ yields
\begin{equation}
 \varphi^1=\frac{\sqrt{2} ||\varphi||^4 V_0 V_1 \pm \sqrt{6 ||\varphi||^8 V_1^2 \left(3 (V_0)^2 - 8 ||\varphi||^6 \right)}}{12 ||\varphi||^6 - 4 (V_0)^2}.\label{junkphi1}
\end{equation}
The constraint on $V_I$'s is
\begin{equation}
 (V_1)^2-(V_5)^2-...-(V_{\tilde{n}})^2>0,\label{constraintsvishypersu2}
\end{equation}
and since $\varphi$'s are real, by (\ref{junkphi1})
\begin{equation}
 (V_0)^2 >\frac{8}{3}||\varphi||^6.\label{voconditionhypsu2}
\end{equation}
The potential evaluated at the critical point is given by
\begin{equation}
 P^{(5)}_{TOT}\arrowvert_{\phi^C}=\frac{\lambda}{4}\left( 6||\varphi||^2 +\frac{\left( V_0 \varphi^1 +\sqrt{2} V_1 ||\varphi||^4 \right)^2}{(\varphi^1)^2 ||\varphi||^4} \right)
\end{equation}
which is positive definite. Now, given a set of $V_I$'s subject to the constraint (\ref{constraintsvishypersu2}), the critical point is determined by $\tilde{n}-4$ equations (\ref{vieqnssu2hyp}) together with equation (\ref{junkphi1}).\footnote{One has to make sure that the equation (\ref{voconditionhypsu2}) holds.} Note that, in some cases, there may be more than one solution because of multi-valuedness of (\ref{junkphi1}). We calculated the Hessian of the potential and showed that it is possible to obtain positive eigenvalues and hence one can have stable deSitter vacua. Because of the lengthiness of the expressions we give a particular example here.

\subparagraph{Example:} Suppose
\begin{equation}
 V_0 = 2,\qquad V_1=1,\qquad V_5=...=V_{\tilde{n}}=0.\nonumber
\end{equation}
There are two critical points, given by
\begin{equation}
 \begin{array}{rclr}
\phi^C_1&:&\varphi^1=-\frac{(\sqrt{33}-1)^{1/3}}{2^{5/6}},& \varphi^5=...=\varphi^{\tilde{n}}=0,\nonumber\\
\phi^C_2&:&\varphi^1=\frac{(\sqrt{33}+1)^{1/3}}{2^{5/6}},& \varphi^5=...=\varphi^{\tilde{n}}=0.
\end{array}\end{equation}
The values of the potential at these critical points read
\begin{equation}\begin{array}{rcl}
 P^{(5)}_{TOT}\arrowvert_{\phi_1^C}&=&\frac{3}{4}\lambda \left( \frac{3}{2} \left(69-11\sqrt{33} \right)\right)^{1/3},\nonumber\\
P^{(5)}_{TOT}\arrowvert_{\phi_2^C}&=&\frac{3}{4}\lambda \left( \frac{3}{2} \left(69+11\sqrt{33} \right)\right)^{1/3}
\end{array}\end{equation}
and the numerical values for the eigenvalues of the Hessian (for $\tilde{n}=6$) are
\begin{equation}\begin{array}{c}
(-0.799 \lambda, -0.799 \lambda, -0.743 \lambda, -0.686 \lambda, -0.686 \lambda,
0.667 \lambda, 1.142 \lambda, 2.991 \lambda, 2.991 \lambda, 29.058 \lambda)\end{array}\nonumber
\end{equation}
at $\phi_1^C$ and
\begin{equation}\begin{array}{c}
( 0.843 \lambda, 1.102 \lambda, 1.102 \lambda, 1.876 \lambda, 2.186 \lambda,
2.186 \lambda, 6.526 \lambda, 7.143 \lambda, 7.143 \lambda, 20.441 \lambda )
\end{array}\nonumber
\end{equation}
at $\phi_2^C$. Hence the second critical point is stable whereas the first one is not.

\subsubsection{$U(1)_R$ symmetry gauging}
\paragraph{Without hypermultiplets:}
See \cite{Gunaydin:1984ak} for a complete analysis for the cases without tensors for all symmetric Jordan theories. Here we will review a specific result, which will be relevant when we will add a hypermultiplet into the theory. As the $U(1)_R$-gauge field, a linear combination $V_I A^I$ of all the vectors in the theory will be taken. Using (\ref{u1r}), the potential is given by
\begin{equation}
 P^{(5)}_{TOT}=\lambda P^{(R)}=-2 \lambda\left(||V||^2 ||\varphi||^2 +\frac{2\sqrt{2}V_0 V_i \varphi^{i}}{||\varphi||^2}\right)\label{pureu1rpot}
\end{equation}
where $i=1,...,\tilde{n}$ and $||V||^2=(V_1)^2 - (V_2)^2 - ...- (V_{\tilde{n}})^2$. The derivatives of this potential are calculated as
\begin{equation}
 \begin{array}{rcl}
 \partial_{\varphi^1} P^{(5)}_{TOT} &=& -2\lambda\left( \varphi^1 \tilde{A} + \frac{\sqrt{2} V_0 V_1}{||\varphi||^2} \right),\\
\partial_{\varphi^a} P^{(5)}_{TOT} &=& -2\lambda\left( -\varphi^a \tilde{A} + \frac{\sqrt{2} V_0 V_a}{||\varphi||^2} \right),\qquad a=2,...,\tilde{n}.\label{u1rderivatives}
\end{array}
\end{equation}
where
\begin{equation}
 \tilde{A}=||V||^2 -\frac{2\sqrt{2} V_0 V_i \varphi^i}{||\varphi||^4}.
\end{equation}

A trivial way of making the derivatives (\ref{u1rderivatives}) vanish is to set $V_i = 0$. This leads to a Minkowski ground state with broken supersymmetry ($P_1\neq 0$) as long as $V_0\neq 0$, i.e. the $U(1)_R$ gauging is nontrivial.

The easiest way to solve the equations nontrivially, after the derivatives are set to zero, is to solve the last equation for $V_{0}$; plug the resulting expression into the other equations; solve the equation before the last equation for $V_{\tilde{n}}$; plug the resulting expression into the remaining equations; solve the last of the remaining equations for $V_{\tilde{n}-1}$; plug the resulting expression into the remaining equations and so forth... At the end one finds
\begin{equation}\begin{array}{rcl}
 \sqrt{2} V_0 \varphi^{1} &=& V_1 ||\varphi||^4\\
\varphi^{1} V_a &=& - \varphi^{a} V_1\label{pureu1rsol}
\end{array}\end{equation}
$V_i$'s satisfy the following constraint
\begin{equation}
 (V_1)^2-(V_2)^2-...-(V_{\tilde{n}})^2 >0.
\end{equation}
By plugging in (\ref{pureu1rsol}) into the potential (\ref{pureu1rpot}), one evaluates its value at the critical point as
\begin{equation}
 P^{(5)}_{TOT}\arrowvert_{\varphi^C}=-6 \lambda (V_1)^2 \frac{||\varphi||^4}{(\varphi^1)^2},\label{pureu1rpotvalue}
\end{equation}
which is negative and therefore corresponds to an AdS critical point. Calculating the Hessian of the potential, one finds that it always has the negative eigenvalue
\begin{equation}
 -4\lambda (V_1)^2 (\varphi^1)^2\frac{2 \left(( \varphi^1)^2+...+(\varphi^{\tilde{n}})^2\right)+\sqrt{||\varphi||^4+16 (\varphi^1)^2 \left((\varphi^2)^2+...+(\varphi^{\tilde{n}})^2\right)}}{(\varphi^1)^4}\nonumber
\end{equation}
for any $\tilde{n}$, hence the critical point is not a minimum. Moreover we found that, up to $\tilde{n}=4$, the eigenvalues of the Hessian are all negative. This means that the critical point is a maximum rather than a minimum.\footnote{See \cite{Gunaydin:1984ak} for the general proof that this is the case for arbitrary $\tilde{n}$.} The unboundedness of the potential from below may lead someone to think that the critical point is unstable. But an analysis of small fluctuations of $\varphi^{\tilde{x}}$ around the critical point shows that, at least perturbatively, instabilities need not occur \cite{Breitenlohner:1982jf}. It was shown in \cite{Townsend:1984iu} and demonstrated in \cite{Gunaydin:1984ak} that a potential of the form (\ref{u1r}) is sufficient to ensure the positivity of the energy and thereby the stability, about the AdS background at a critical point. To have supersymmetry at the critical point one needs to have $\langle P^{i}\rangle = 0$. We calculated $P_{i} = -\sqrt{\frac{3}{2}} P_{,i}=-\sqrt{\frac{3}{2}} (V_I h^I)_{,i}$ as
\begin{equation}
 \begin{array}{rcl}
  P_1 &=&\displaystyle  \frac{\sqrt{2} V_0 \varphi^1}{||\varphi||^4}-V_1\\[8pt]
P_a &=&\displaystyle  -\frac{\sqrt{2} V_0 \varphi^a}{||\varphi||^4}-V_a\nonumber
 \end{array}
\end{equation}
and eqns. (\ref{pureu1rsol}) assure that these quantities vanish and hence the critical point is supersymmetric.

\paragraph{With a universal hypermultiplet:}
The total potential is of the form $P_{TOT}^{(5)}=P^{(R)}+P^{(H)}$. The most general way of doing simultaneous $U(1)_R$ gauging together with $U(1)$ gauging of the hypermultiplet isometry is done by selecting a linear combination of compact Killing vectors from (\ref{Tvex}). One can easily see that at the base point $q^c = \{V=1, \sigma=\theta=\tau=0\}$ of the hyperscalar manifold all these compact generators vanish. Therefore one has $\mathcal{N}^{iA}=0$ and as a consequence \cite{smet}
\begin{equation}
 P^{(H)}\arrowvert_{q^c}=\frac{\partial P^{(H)}}{\partial \varphi^I}\arrowvert_{q^c}=\frac{\partial P^{(H)}}{\partial q}\arrowvert_{q^c}= 0.\label{u1saddle}
\end{equation}
On the other hand, $P^{(R)}$ of (\ref{pr_h}) is of the form $P^{(R)}\sim f(\varphi) g(q)$, where $g(q)=\vec{P}_I \cdot \vec{P}_J (q) \delta^{IJ}$ for the generic family. $g(q)$ has an extremum point at the base point of the hyperscalar manifold (i.e. $\frac{dg}{dq}\arrowvert_{q^c}=0$). This leads to
\begin{equation}
 \frac{\partial P^{(R)}}{\partial q}\arrowvert_{q^c}=\frac{\partial P^{(5)}_{TOT}}{\partial q}\arrowvert_{q^c}=\frac{\partial^2 P^{(5)}_{TOT}}{\partial \varphi \partial q}\arrowvert_{q^c}=0\nonumber
\end{equation}
and hence the Hessian is in block diagonal form. We already showed that the pure $U(1)_R$ gauging lead to at least one negative eigenvalue of the Hessian. The fact that $g(q) \geq 0$ makes it impossible to convert the non-minimum critical points that correspond to the upper block of the Hessian ($\frac{\partial^2 P^{(5)}_{TOT}}{(\partial \varphi)^2}$) to minimum points of the potential or change its sign at the critical point. Therefore a $U(1)_H$ gauging will not change the nature of an existing critical point.

However, one has to check what the noncompact generators would do for which (\ref{u1saddle}) does not hold. 
\subparagraph{$U(1)_R \times SO(1,1)_H$ gauging:}
For the $SO(1,1)$ gauging, a linear combination $W_I A^I$ of all the vectors of the theory will be taken as the gauge field. The $U(1)_R$ gauge field must be orthogonal to the $SO(1,1)$ gauge field. This leads to the condition
\begin{equation}
 V_I W_I = 0.\label{vwconstr}
\end{equation}
The potential is again given by
\begin{equation}
 P^{(5)}_{TOT}=\lambda (P^{(R)}+2\mathcal{N}_{iA}\mathcal{N}^{iA})
\end{equation}
where this time $\mathcal{N}^{iA}=\frac{\sqrt{6}}{4} (V_I h^I Y^a T_a^X+ W_I h^I T_4^X) f^{iA}_X$ where $Y^a T_a^X$, with $a=1,2,3$, defines the linear combination of compact Killing vectors to be used; the $SO(1,1)$ coupling constant is absorbed in $W_I$'s and the $P^{(R)}$ term is
\begin{equation}
 P^{(R)}=-4 C^{IJK} V_I V_J h_K (Y^a \vec{P}_a)\cdot (Y^b \vec{P}_b).
\end{equation}
The first derivatives of the potential vanish by using (\ref{pureu1rsol}) and by setting 
\begin{equation}
 W_1 =\frac{-\sqrt{2} W_0 - 2 W_b \varphi^b ||\varphi||^2}{2 \varphi^1 ||\varphi||^2},\qquad b=2,...,\tilde{n}.\label{u1rhypso11}
\end{equation}
Plugging in everything into the potential, one finds
\begin{equation}
 P^{(5)}_{TOT}\arrowvert_{\phi^C}=-\frac{3}{2} \lambda \left(\frac{V_1}{\varphi^1}\right)^2 ||\varphi||^4 (Y^a Y^a)
\end{equation}
which is manifestly negative and therefore this corresponds to an AdS ground state. Note that all the $\varphi^{\tilde{x}}$'s in this equation are fixed by (\ref{pureu1rsol}).
The stability is checked by calculating the eigenvalues of the Hessian of the poteintial. The calculation is tedious and although we were not able to prove generally, all the gauge field combinations subject to (\ref{pureu1rsol}), (\ref{vwconstr}) and (\ref{u1rhypso11}) that we tried lead to negative eigenvalues for the Hessian and hence the corresponding critical points were not minima. However the potential is in the form suggested by \cite{Townsend:1984iu} plus a term that is quadratic in $h^I$ hence it is possible to obtain stable AdS vacua with proper choices of $V_I$ and $W_I$, provided that the eigenvalues of the Hessian that belong to the hyperscalar sector are positive.
\subsection{YMESGT with compact ($SO(2)$) gauging, coupled to tensor fields}
The calculation was done in \cite{Gunaydin:Vacua} for $\tilde{n}=3$. Let us trivially generalize their results to arbitrary $\tilde{n}\geq 3$. The $SO(2)$ subgroup of the isometry group of the scalar manifold acts nontrivially on the vector fields $A^2_\mu$ and $A^3_\mu$. Hence these vector fields must be dualized to antisymmetric tensor fields. The index $\tilde{I}$ is decomposed as
\begin{equation}
 \tilde{I}=(I,M)\nonumber
\end{equation}
 with $I,J,K=0,1,4,...,\tilde{n}$ and $M,N,P=2,3$. The fact that the only nonzero $C_{IMN}$ are $C_{0MN}$ for the theory at hand requires $A_\mu^0$ to be the $SO(2)$ gauge field because of $\Lambda^M_{IN}\sim \Omega^{MP} C_{IPN}$ (c.f equation (\ref{pt})). All the other $A_\mu^I$ with $I\neq 0$ are spectator vector fields with respect to the $SO(2)$ gauging. The potential term (\ref{pt}) that comes from the tensor coupling is found to be (taking $\Omega^{23}=-\Omega^{32}=-1$)
\begin{equation}
 P^{(T)}=\frac{1}{8} \frac{\left[ (\varphi^2)^2 + (\varphi^3)^2\right]}{||\varphi||^6}\label{so2pt}
\end{equation}
For the function $W_{\tilde{x}}$ that enters the supersymmetry transformation laws of the fermions, one obtains
\begin{equation}
 \begin{array}{rcl}
  W_1=W_4=&...&=W_{\tilde{n}}=0,\\
W_2&=&\frac{\varphi^3}{4||\varphi||^4},\\
W_3&=&-\frac{\varphi^2}{4||\varphi||^4},
 \end{array}
\end{equation}
so one must have $\varphi^2_C=\varphi^3_C=0$ to preserve supersymmetry.
\subsubsection{No $R$-symmetry gauging}
\paragraph{Without hypermultiplets:}
Taking the derivative of the total potential $P^{(5)}_{TOT}=P^{(T)}$ with respect to $\varphi^{\tilde{x}}$, one finds
\begin{equation}
 \begin{array}{rcll}
 \partial_{\varphi^1} P^{(5)}_{TOT} &=&\displaystyle  -\frac{3}{4} \frac{\left[ (\varphi^2)^2 + (\varphi^3)^2\right]}{||\varphi||^8} \varphi^1,\\[6pt]
\partial_{\varphi^a} P^{(5)}_{TOT} &=&  A \varphi^a,&a=2,3;\\[2pt]
\partial_{\varphi^b} P^{(5)}_{TOT} &=&\displaystyle  \frac{3}{4} \frac{\left[ (\varphi^2)^2 + (\varphi^3)^2\right]}{||\varphi||^8} \varphi^b,\quad&b=4,...,\tilde{n}
\end{array}\nonumber
\end{equation}
where
\begin{equation}
 A=\frac{1}{4} \frac{||\varphi||^2 +3 \left[ (\varphi^2)^2 + (\varphi^3)^2\right]}{||\varphi||^8} > 0 .\nonumber
\end{equation}
$\partial_{\varphi^a} P^{(5)}_{TOT}=0$ then implies $\varphi_C^2=\varphi_C^3=0$ (which then also implies $\partial_{\varphi^{\tilde{x}}} P^{(5)}_{TOT} \arrowvert_{\varphi^C}= 0, \forall \tilde{x}$). But then $P^{(5)}_{TOT}\arrowvert_{\varphi^C}=0$ and we have a $\tilde{n}-2$ parameter family of supersymmetric Minkowski ground states, given by $\langle \varphi^2\rangle = \langle \varphi^3\rangle = 0$ and arbitrary $\langle\varphi^{d}\rangle, \quad d=1,4,5,...,\tilde{n}$.
\paragraph{With a universal hypermultiplet:}
The compact generators of (\ref{Tvex}) vanish at the base point of the hyperscalar manifold. Hence a $U(1)$ gauging of the hyper isometries will not introduce a non-Minkowski ground state.
\subparagraph{$SO(2)\times SO(1,1)_H$ gauging:}
The $SO(1,1)$ gauge field is chosen as a linear combination of all vector fields that are not dualized to tensor fields. The total potential therefore is
\begin{equation}
 P^{(5)}_{TOT} = P^{(T)} + \kappa P^{(H)}\nonumber
\end{equation}
where $P^{(T)}$ was given in (\ref{so2pt}) and $P^{(H)}=2\mathcal{N}_{iA}\mathcal{N}^{iA}$ where
\begin{equation}
 \mathcal{N}^{iA} = V_e h^e T_4^X f_X^{iA},\qquad e=0,1,4,5,...,\tilde{n}
\end{equation}
with $T_4^X$ given in (\ref{Tvex}). At the base point of the hyperscalar manifold the derivatives of the total potential with respect to $q^X$ vanish. One can calculate the $\varphi^a$-derivatives as
\begin{equation}
 \partial_{\varphi^a} P^{(5)}_{TOT}\arrowvert_{q^C}=\frac{\varphi^a \tilde{B}}{4 ||\varphi||^8}\label{so2hypso11ppat}
\end{equation}
where 
\begin{equation}\begin{array}{rcl}
 \tilde{B}&=&||\varphi||^2+3 \left[ (\varphi^2)^2+(\varphi^3)^2\right]\\
&&+4\kappa V_0 ||\varphi||^2 \left\{ V_0 +\sqrt{2} \left(V_1\varphi^1+V_4\varphi^4+...+V_{\tilde{n}}\varphi^{\tilde{n}}\right)||\varphi||^2\right\}.\nonumber
\end{array}\end{equation}
There are two possible ways to make (\ref{so2hypso11ppat}) vanish.
\subparagraph{\underline{Case 1: $\tilde{B}=0$}}
One can solve the equation $\tilde{B}=0$ for $V_1$ and plug that into\newline $\partial_{\varphi^b} P^{(5)}_{TOT}\arrowvert_{q^C}=0$ to get
\begin{equation}
 -\frac{2V_0 \varphi^b +\sqrt{2} V_b ||\varphi||^2 \left\{||\varphi||^2+3\left[(\varphi^2)^2+(\varphi^3)^2\right]\right\}}{8 V_0 ||\varphi||^6}=0\nonumber
\end{equation}
Solving this for $V_b$ and plugging the resulting expression together with the $\tilde{B}=0$ equation into $\partial_{\varphi^1} P^{(5)}_{TOT}\arrowvert_{q^C}$, one finds
\begin{equation}
 \partial_{\varphi^1} P^{(5)}_{TOT}\arrowvert_{q^C}\rightarrow\text{{\small $ \frac{9\left[(\varphi^2)^2+(\varphi^3)^2\right] +2 (3+10 \kappa (V_0)^2) \left[(\varphi^2)^2+(\varphi^3)^2\right] ||\varphi||^2+(1+12\kappa (V_0)^2)||\varphi||^4 }{32\kappa (V_0)^2 \varphi^1 ||\varphi||^8} $ } }\nonumber
\end{equation}
which cannot be brought to zero. Hence there is no solution for this case.
\subparagraph{\underline{Case 2: $\tilde{B}\neq 0$}}
One has to have $\varphi^a=0$. Applying this to the remaining first derivative equations, all there are left to solve are the following expressions
\begin{equation}
 \begin{array}{rcl}
 \displaystyle  \frac{\kappa}{||\varphi||^2}\left\{\frac{V_1}{\sqrt{2}}-\frac{V_0 \varphi^1}{||\varphi||^4}\right\}\tilde{C}&=&0,\\[8pt]
 \displaystyle  \frac{\kappa}{||\varphi||^2}\left\{\frac{V_b}{\sqrt{2}}+\frac{V_0 \varphi^b}{||\varphi||^4}\right\}\tilde{C}&=&0
 \end{array}
\end{equation}
 with
\begin{equation}
 \tilde{C} = V_0 +\sqrt{2} \left(V_1 \varphi^1+V_4 \varphi^4+V_5 \varphi^5+...+V_{\tilde{n}} \varphi^{\tilde{n}}\right)||\varphi||^2.
\end{equation}
It is possible to set both expressions to zero by letting $\tilde{C}=0$, but this will make the potential vanish at the critical point. Setting $\kappa=0$ will turn off the hyper-gauging. Instead letting
\begin{equation}
 \frac{\sqrt{2} V_0}{||\varphi||^4}=\frac{V_1}{\varphi^1}=-\frac{V_4}{\varphi^4}=-\frac{V_5}{\varphi^5}=...=-\frac{V_{\tilde{n}}}{\varphi^{\tilde{n}}}\label{so2hypso11cond}
\end{equation}
will make them vanish and the value of the potential at the critical point becomes
\begin{equation}
 P^{(5)}_{TOT} \arrowvert_{\phi^C}=\frac{9\kappa (V_0)^2}{4||\varphi||^4}.
\end{equation}
 which is positive definite and hence the critical point is a deSitter ground state. The equations (\ref{so2hypso11cond}) set the restrictions on choosing $V_I$ as
\begin{equation}\begin{array}{c}
 (V_1)^2-(V_4)^2-(V_5)^2-...-(V_{\tilde{n}})^2>0\\
V_0\neq 0\nonumber
\end{array}\end{equation}

Given a set of $V_I$'s subject to these constraints, the coordinates of the critical point is totally determined by (\ref{so2hypso11cond}). The non-negativeness of the eigenvalues of the Hessian of the potential assures the stability of the vacuum. For the special case $V_4=...=V_{\tilde{n}}=0$ the Hessian is calculated as
\begin{equation}\begin{array}{rl}
 \partial\partial P^{(5)}_{TOT}\arrowvert_{\phi^C}=\text{diag}&\displaystyle (\frac{9\kappa(V_0)^2}{(\varphi^1)^6},\frac{1+12\kappa(V_0)^2}{4(\varphi^1)^6},\frac{1+12\kappa(V_0)^2}{4(\varphi^1)^6},\overbrace{\frac{3\kappa(V_0)^2}{(\varphi^1)^6},...,\frac{3\kappa(V_0)^2}{(\varphi^1)^6}}^{(\tilde{n}-3) \text{ times}},\\[8pt]
&\displaystyle 0,\frac{9\kappa(V_0)^2}{(2\varphi^1)^4},\frac{9\kappa(V_0)^2}{(4\varphi^1)^4},\frac{9\kappa(V_0)^2}{(4\varphi^1)^4}).\nonumber
\end{array}\end{equation}
and therefore the ground state is stable. For the more general case the Hessian is not diagonal, but we were able to show that the eigenvalues of the Hessian are non-negative up to at least $\tilde{n}=6$.  
\subsubsection{$SU(2)_R$ symmetry gauging}
\paragraph{Without hypermultiplets:}
The gauge group is $SO(2)\times SU(2)_R$. For such a gauging one needs at least $\tilde{n}\geq 6$. Choosing $A^4_\mu,A^5_\mu$ and $A^6_\mu$ as the $SU(2)_R$ gauge fields one finds \cite{Gunaydin:2000ph}
\begin{equation}
 P^{(5)}_{TOT} = P^{(T)} + \lambda P^{(R)}\nonumber
\end{equation}
 with
\begin{equation}
 P^{(R)}=6 ||\varphi||^2\nonumber
\end{equation}
and $P^{(T)}$ given in (\ref{so2pt}). It is easy to verify that the total potential does not have any non-Minkowskian ground states. In particular, in order to set the first derivatives to zero, one must have $\varphi_C^2=\varphi_C^3=\lambda=0$ which means the $SU(2)_R$ gauging is turned off and this case was already covered in the previous section.
\paragraph{With a universal hypermultiplet:}
Inclusion of a hypermultiplet in the theory will change the potential to
\begin{equation}
 P^{(5)}_{TOT} = P^{(T)} + \lambda (P^{(R)} + P^{(H)})\nonumber
\end{equation}
with now
\begin{equation}\begin{array}{rcl}
P^{(R)}&=&-4 C^{IJK} \vec{P}_I\cdot \vec{P}_J h_K,\\
P^{(H)}&=&\frac{3}{4} h^I h^J K_I^X K_J^Y g_{XY}
\end{array}
\end{equation}
where $K_I^X$ are defined as
\begin{equation}
K_4^X = T_1^X \qquad K_5^X = T_2^X \qquad K_6^X = T_3^X
\end{equation}
 and $\vec{P}_I$ are defined accordingly. Remember that $K_I^X=0$ for compact generators at the base point of the hyperscalar manifold and therefore one has $P^{(H)}\arrowvert_{q^C}=0$. It is easy to see that this case is very similar to the case before adding the hypermultiplet and the only possibility is to have Minkowski vacuum. An additional $U(1)_H$ gauging will not change the situation but let us see what would the $SO(1,1)_H$ gauging do.
\subparagraph{$SO(2)\times SU(2)_R \times SO(1,1)_H$ gauging:}
The $SO(1,1)$ gauge field is chosen as the linear combination $V_a A_\mu^a,\quad a=0,1,7,8,...,\tilde{n}$. The total potential is
\begin{equation}
 P^{(5)}_{TOT} = P^{(T)} + \lambda (P^{(R)} + P^{(H)})\nonumber
\end{equation}
where $P^{(T)}$ and $P^{(R)}$ are as given as in the last case and $P^{(H)}=2\mathcal{N}_{iA} \mathcal{N}^{iA}$ is modified with
\begin{equation}
 \mathcal{N}^{iA}=\frac{\sqrt{6}}{4} (h^I K_I^X +h^a V_a T^X_4) f_X^{iA}.
\end{equation}
The calculations for finding the critical points is overly complicated and the expressions are lenghty. Here, we will show a particular example where a stable deSitter vacuum is found. The first derivatives of the total potential vanish at\footnote{There are other critical points where $\varphi^e (e=2,...,\tilde{n})$ are not all zero, but we found that they are unstable.} $V=1, \sigma=\theta=\tau=\varphi^2=...=\varphi^8=0$ except the $\varphi^1$-derivative 
\begin{equation}
  \partial_{\varphi^1} P^{(5)}_{TOT}\arrowvert_{\varphi^C}= 9\lambda \left\{ (1+ (V_1)^2)\varphi^1 -\frac{ (V_0)^2}{(\varphi^1)^5}-\frac{V_0 V_1}{\sqrt{2} (\varphi^1)^2}\right\}.\nonumber
\end{equation}
Setting this to zero determines the $\varphi^1$-coordinates of the critical points as a function of $V_0$ and $V_1$
\begin{equation}
 \varphi^1=\frac{1}{\sqrt{2}}\left( \frac{ V_0 V_1 \pm \sqrt{ (V_0)^2(8+9(V_1)^2)}}{1+ (V_1)^2}\right)^{1/3}\label{sfgnbf}.
\end{equation}
The values of the total potential at these critical points are
\begin{equation}
 P^{(5)}_{TOT}\arrowvert_{\phi^C}=\frac{27\lambda V_0 \left(3 V_0 (\varphi^1)^3 \pm\sqrt{9(V_0)^2 (\varphi^1)^6-8(\varphi^1)^{12}}\right)}{8(\varphi^1)^7}.\nonumber
\end{equation}
where $\varphi^1$ was given in (\ref{sfgnbf}). The values of the potential are positive definite and therefore the critical points correspond to deSitter vacua. 

\subparagraph{Example:} In particular, we look at the $\tilde{n}=8$ theory by taking $V_0=1$ and $V_1=4$. There is a critical point located at $\varphi^1=\left(\frac{2+\sqrt{38}}{17\sqrt{2}}\right)^{1/3}$. The value of the potential at this point is $\frac{27}{4} (937 +152\sqrt{38})^{1/3} \lambda$. With these choices one can calculate the eigenvalues of the Hessian of the potential numerically as
\begin{equation}\begin{array}{l}
 \{1.095 \lambda, 19.574 \lambda, 19.574 \lambda, 127.337 \lambda, 218.959 \lambda, 218.959 \lambda, 254.777\lambda\\
 284.796 \lambda, 284.796 \lambda, 693.122 \lambda, 2.168+218.959 \lambda, 2.168+218.959 \lambda\}\nonumber
\end{array}\end{equation}
This shows that the critical point is stable.
\subsubsection{$U(1)_R$ symmetry gauging\label{genericso2u1r}}
\paragraph{Without hypermultiplets:}
The gauge group is $SO(2)\times U(1)_R$. For such a gauging one needs at least $\tilde{n}\geq 3$. A linear combination $A_\mu [U(1)_R]=V_I A^I_\mu$ of vector fields will be used as the $U(1)_R$ gauge field. The total scalar potential in this case is
\begin{equation}
 P^{(5)}_{TOT}=P^{(T)}+\lambda P^{(R)}
\end{equation}
with
\begin{equation}
 P^{(R)}=-2 |V|^2 ||\varphi||^2 -4\sqrt{2} V_0 \frac{V_i \varphi^i}{||\varphi||^2}
\end{equation}
where $i=1,4,5,...,\tilde{n},\quad |V|^2 = (V_1)^2 - (V_4)^2-...-(V_{\tilde{n}})^2$ and $P^{(T)}$ given in (\ref{so2pt}). The first derivatives of the potential
\begin{equation}
 \begin{array}{rcl}
  \partial_{\varphi^1} P^{(5)}_{TOT}&=& - (D\varphi^1+4\lambda C V_1),\\
\partial_{\varphi^a} P^{(5)}_{TOT}&=& \varphi^a (D+\frac{1}{4||\varphi||^6}),\qquad a=2,3;\\
\partial_{\varphi^b} P^{(5)}_{TOT}&=&  (D\varphi^b-4\lambda C V_b),\qquad b=4,...,\tilde{n}
 \end{array}
\end{equation}
must simultaneously vanish at the critical point(s). Here we defined
\begin{equation}
 \begin{array}{rcl}
  C&=&\displaystyle \frac{\sqrt{2} V_0}{||\varphi||^2}\\[6pt]
D&=&\displaystyle \frac{6 P^{(T)}}{||\varphi||^2}+4\lambda \left(|V|^2 - \frac{2\sqrt{2} V_0 V_i \varphi^i}{||\varphi||^4}\right).
 \end{array}
\nonumber
\end{equation}
There are two possibilities to set the second equation to zero.
\subparagraph{\underline{Case 1: $\varphi^a=0$}}
This means that $P^{(T)}\arrowvert_{\varphi^C}=\partial_{\varphi^{\tilde{x}}} P^{(T)}\arrowvert_{\varphi^C}=0$ and consequently \newline\mbox{$\partial_{\varphi^{\tilde{x}}} P^{(R)}\arrowvert_{\varphi^C}=0$}. Thus we are dealing with simultaneous critical points of the individual potentials $P^{(T)}$ and $P^{(R)}$. These have already been discussed above. In particular, the coordinates of the critical points are entirely determined by (\ref{pureu1rsol}, with $a=4,...,\tilde{n}$) and the potential corresponds to a supersymmetric Anti-deSitter vacuum with the value given in (\ref{pureu1rpotvalue}). Also, it is possible to have a Minkowski ground state with broken supersymmetry by letting all $V_I$ vanish, except $V_0$.
\subparagraph{\underline{Case 2: $\varphi^a\neq 0$}}
In this case one must have
\begin{equation}
 D=-\frac{1}{4||\varphi||^6}.\label{so2d1}
\end{equation}
The first and the last equations tell that
\begin{equation}
 -\frac{V_1}{\varphi^1}=\frac{V_b}{\varphi^b}=\frac{D}{4\lambda C}\label{so2fracv}
\end{equation}
which means $|V|^2>0$ and hence
\begin{equation}
 D=-\frac{4\lambda V_1}{\varphi^1} C=-\frac{4\sqrt{2} \lambda V_0 V_1}{\varphi^1 ||\varphi||^2}.\label{so2d2}
\end{equation}
This leads to
\begin{equation}
 \varphi^1=16\sqrt{2}\lambda V_0 V_1 ||\varphi||^4\label{so2varphi1}
\end{equation}

Plugging (\ref{so2d2}) and (\ref{so2varphi1}) into (\ref{so2d1}) one arrives at
\begin{equation}
 \frac{1}{2||\varphi||^6}=384 \lambda^2 (V_0)^2 (V_1)^2 +4\lambda |V|^2 (1-64\lambda (V_0)^2).\label{so2arrive}
\end{equation}
Using $|V|^2>0$ together with
\begin{equation}
 (32\lambda (V_0)^2-1)(V_1)^2> (64\lambda(V_0)^2-1)(V_b V_b)\nonumber
\end{equation}
which can be derived from (\ref{so2arrive}), one obtains the condition
\begin{equation}
 32\lambda(V_0)^2>1.
\end{equation}
If $V_0$ is chosen big enough to satisfy this, new non-trivial critical points exist. Eq. (\ref{so2arrive}) fixes $||\varphi_C||^2$ so that eq.'s (\ref{so2fracv}) and (\ref{so2varphi1}) fix $\varphi_C^1$ and $\varphi_C^b$. This in turn fixes $[(\varphi_C^2)^2+(\varphi_C^3)^2]$ but not $\varphi_C^2$ and $\varphi_C^2$ individually. Therefore we have a one parameter family of critical points. The value of the potential at the critical points is
\begin{equation}
 P^{(5)}_{TOT}\arrowvert_{\varphi^C}=-32\lambda^2 (V_0)^2 (V_1)^2 ||\varphi||^2 -\lambda ||\varphi||^2 |V|^2 (3+64\lambda (V_0)^2)
\end{equation}
which corresponds to a non-supersymmetic Anti-deSitter solution. This result agrees with \cite{Gunaydin:Vacua} in the $\tilde{n}=3$ limit\footnote{$P^{(5)}_{TOT,\tilde{n}=3}\arrowvert_{\varphi^C}=-\frac{3}{8}\frac{1}{||\varphi||^4}$}. As pointed out in that work, these critical points are saddle points of the total potential. The potential is in the form suggested in \cite{Townsend:1984iu} plus the semi-positive definite paraboloidlike $P^{(T)}$ term. This tells us that the ground state is stable.
\paragraph{With a universal hypermultiplet:}
Inclusion of a hypermultiplet in the theory changes the total potential to
\begin{equation}
  P^{(5)}_{TOT}=P^{(T)} +\lambda (P^{(R)} +2\mathcal{N}_{iA} \mathcal{N}^{iA})
\end{equation}
with now
\begin{equation}
 \begin{array}{rcl}
  P^{(R)}&=&-4 C^{IJK} \vec{P}_I \cdot \vec{P}_J h_K\\
\mathcal{N}^{iA}&=&\frac{\sqrt{6}}{4} (V_I h^I Y^d T_d^X) f_X^{iA}
 \end{array}
\end{equation}
where $Y^d T_d^X$ with $d=1,2,3,8$ defines the linear combination of compact Killing vectors to be used. Remember that $K_I^X=0$ for compact Killing vectors at the base point of the hyperscalar manifold and therefore one has $P^{(H)}\arrowvert_{q^C}=0$. In the last subsection we showed that a $U(1)_H$ gauging will not change the nature of the existing critical points in the theory and hence the critical points are saddle points in this case too.
\subparagraph{$SO(2)\times U(1)_R \times SO(1,1)_H$ gauging:}
This is very similar to the previous case. The only difference is
\begin{equation}
 \mathcal{N}^{iA}=\frac{\sqrt{6}}{4}(V_I h^I Y^d T_d^X+W_I h^I T_4^X) f_X^{iA}.
\end{equation}
The linear combination $W_I h^I$ of the vector fields is used as the $SO(1,1)$ gauge field. The $SO(1,1)$ coupling constant is absorbed in $W_I$'s and the fact that the $U(1)_R$ gauge vector field must be orthogonal to the $SO(1,1)_H$ gauge field tells the orthogonality condition
\begin{equation}
 V_I W_I = 0.
\end{equation}
The only nontrivial way to set the first derivatives of the potential to zero we found was done by using (\ref{u1rhypso11} with $b=4,..,\tilde{n}$), (\ref{so2d1}) and (\ref{so2fracv}) but this means that $\partial_{\varphi^{\tilde{x}}} (P^{(T)} + \lambda P^{(R)})$ and $\partial_{\varphi^{\tilde{x}}} P^{(H)}$ must vanish separately. Thus we are dealing with simultaneous critical points of the individual potentials $P^{(T)} + \lambda P^{(R)}$ and $P^{(H)}$ which have already been discussed above. In particular, the value of the potential at the one parameter family of critical points becomes
\begin{equation}
P^{(5)}_{TOT}\arrowvert_{\phi^C}=-32\tilde{\lambda}^2 (V_0)^2 (V_1)^2 ||\varphi||^2 -\tilde{\lambda} ||\varphi||^2 |V|^2 (3+64\lambda (V_0)^2)
\end{equation}
where $\tilde{\lambda}=\frac{\lambda}{4}[(Y^2)^2+(Y^3)^2+(Y^4)^2]$ and it corresponds to an Anti-deSitter ground state, which is of the same form (up to a positive rescaling of $\lambda$) as before the hypermultiplet was added to the theory We expect that it may be possible to obtain stable vacuum with proper choices of the gauge parameters $V_I$ and $W_I$, provided that the eigenvalues of the Hessian that belong to the hyperscalar sector are positive.
\subsection{YMESGT with non-compact ($SO(1,1)$) gauging, coupled to tensor fields}
The calculation was done in \cite{Gunaydin:Vacua} for $\tilde{n}=3$. Let us trivially generalize their results to arbitrary $\tilde{n}\geq 2$. The $SO(1,1)$ subgroup of the isometry group of the scalar manifold acts nontrivially on the vector fields $A^1_\mu$ and $A^2_\mu$. Hence these vector fields must be dualized to antisymmetric tensor fields. The index $\tilde{I}$ is decomposed as
\begin{equation}
 \tilde{I}=(I,M)\nonumber
\end{equation}
 with $I,J,K=0,3,4,...,\tilde{n}$ and $M,N,P=1,2$. The fact that the only nonzero $C_{IMN}$ are $C_{0MN}$ for the theory at hand requires $A_\mu^0$ to be the $SO(1,1)$ gauge field because of $\Lambda^M_{IN}\sim \Omega^{MP} C_{IPN}$ (c.f equation (\ref{pt})). All the other $A_\mu^I$ with $I\neq 0$ are spectator vector fields with respect to the $SO(1,1)$ gauging. The potential term (\ref{pt}) that comes from the tensor coupling is found to be (taking $\Omega^{23}=-\Omega^{32}=-1$)
\begin{equation}
 P^{(T)}=\frac{1}{8} \frac{\left[ (\varphi^1)^2 - (\varphi^2)^2\right]}{||\varphi||^6}\label{so11pt}
\end{equation}
For the function $W_{\tilde{x}}$ that enters the supersymmetry transformation laws of the fermions, one obtains
\begin{equation}
 \begin{array}{rcl}
  W_3=W_4=&...&=W_{\tilde{n}}=0,\\
W_1&=&-\frac{\varphi^2}{4||\varphi||^4},\\
W_2&=&\frac{\varphi^1}{4||\varphi||^4}.
 \end{array}
\end{equation}
Since $W_2$ can never vanish, there can be no $\mathcal{N}=2$ supersymmetric critical point.
\subsubsection{No $R$-symmetry gauging}
\paragraph{Without hypermultiplets:}
Taking the derivative of the total potential $P^{(5)}_{TOT}=P^{(T)}$ with respect to $\varphi^{\tilde{x}}$, one finds
\begin{equation}
 \begin{array}{rcll}
 \partial_{\varphi^1} P^{(5)}_{TOT} &=& B \varphi^1,\\
\partial_{\varphi^2} P^{(5)}_{TOT} &=& - B \varphi^2,\\
\partial_{\varphi^b} P^{(5)}_{TOT} &=& -B \varphi^b +\frac{\varphi^b}{4||\varphi||^6},\quad&b=3,...,\tilde{n}
\end{array}\nonumber
\end{equation}
where
\begin{equation}
 B=-\frac{3}{4}\frac{(\varphi^1)^2-(\varphi^2)^2}{||\varphi||^8}+\frac{1}{4||\varphi||^6}<0 .\label{so11B}
\end{equation}
Since $\partial_{\varphi^1} P^{(5)}_{TOT}$ cannot be brought to zero there are no critical points.
\paragraph{With a universal hypermultiplet:}
The compact generators of (\ref{Tvex}) vanish at the base point of the hyperscalar manifold. Hence a $U(1)$ gauging of the hyper isometries will not introduce critical points.
\subparagraph{$SO(1,1)\times SO(1,1)_H$ gauging:}
The $SO(1,1)_H$ gauge field is chosen as a linear combination of all vector fields that are not dualized to tensor fields. The total potential therefore is
\begin{equation}
 P^{(5)}_{TOT} = P^{(T)} + \kappa P^{(H)}\nonumber
\end{equation}
where $P^{(T)}$ was given in (\ref{so2pt}) and $P^{(H)}=2\mathcal{N}_{iA}\mathcal{N}^{iA}$ where
\begin{equation}
 \mathcal{N}^{iA} = V_e h^e T_4^X f_X^{iA},\qquad e=0,3,4,5,...,\tilde{n}\nonumber
\end{equation}
with $T_4^X$ given in (\ref{Tvex}). At the base point of the hyperscalar manifold the $q$-derivatives of the total potential vanish. One can calculate the $\varphi^1$-derivative as
\begin{equation}
 \partial_{\varphi^1} P^{(5)}_{TOT}\arrowvert_{q^C}=-\frac{\varphi^1 \left(||\varphi||^2-3  [(\varphi^1)^2-(\varphi^2)^2]  +4\kappa V_0 ||\varphi||^2 \left(V_0+\sqrt{2} V_i \varphi^i ||\varphi||^2\right)\right)}{4||\varphi||^8}\nonumber
\end{equation}
where $i=3,...,\tilde{n}$. Setting this expression to zero and solving for $V_{\tilde{n}}$ and plugging the resulting expression into the $\partial_{\varphi^j} P^{(5)}_{TOT}\arrowvert_{q^C}=0,\quad j=2,...,\tilde{n}-1$ equations gives
\begin{equation}
 \begin{array}{rcl}
 \partial_{\varphi^2} P^{(5)}_{TOT}\arrowvert_{\phi^C}&=&0,\\
 \partial_{\varphi^k} P^{(5)}_{TOT}\arrowvert_{\phi^C}&=&\frac{2 V_0 \varphi^k +\sqrt{2} V_k \left(||\varphi||^2 \left(||\varphi||^2 -3 \left[(\varphi^1)^2-(\varphi^2)^2\right]\right)\right)}{8 V_0 ||\varphi||^6}=0;\qquad k=3,...,\tilde{n}-1.\nonumber
 \end{array}
\end{equation}
Solving the equations in the second line for $V_k$ and plugging in everything into the $\varphi^{\tilde{n}}$-derivative of the potential gives
\begin{equation}
 \partial_{\phi^{\tilde{n}}} P^{(5)}_{TOT}\arrowvert_{\varphi^C}=\frac{\left(3\left[(\varphi^1)^2-(\varphi^2)^2\right]-||\varphi||^2\right)^2 +4 ||\varphi||^2 \left(5\left[(\varphi^1)^2-(\varphi^2)^2\right]-3||\varphi||^2\right)\kappa (V_0)^2}{32\kappa (V_0)^2 \varphi^{\tilde{n}} ||\varphi||^8}\nonumber
\end{equation}
and this cannot be brought to zero. Therefore there are no critical points for this type of gauging either.

\subsubsection{$SU(2)_R$ symmetry gauging}
\paragraph{Without hypermultiplets:}
The gauge group is $SO(1,1)\times SU(2)_R$. For such a gauging one needs at least $\tilde{n}\geq 5$. Choosing $A_\mu^3,A_\mu^4,A_\mu^5$ as the $SU(2)_R$ gauge fields one finds
\begin{equation}
 P^{(5)}_{TOT}=P^{(T)}+\lambda P^{(R)}\nonumber
\end{equation}
with
\begin{equation}
 P^{(R)}=6||\varphi||^2
\end{equation}
and $P^{(T)}$ given in (\ref{so11pt}). Taking the derivative of the total potential with respect to $\varphi^{\tilde{x}}$ one finds
\begin{equation}
\begin{array}{rcl}
 \partial_{\varphi^1} P^{(5)}_{TOT}&=& (B+12\lambda)\varphi^1\\
 \partial_{\varphi^2} P^{(5)}_{TOT}&=& -(B+12\lambda)\varphi^2\\
 \partial_{\varphi^b} P^{(5)}_{TOT}&=& -(B+12\lambda)\varphi^b +\frac{\varphi^b}{4||\varphi||^6},\qquad b=3,...,\tilde{n}
\end{array}
\end{equation}
with $B$ defined in (\ref{so11B}). Setting the first equation to zero means
\begin{equation}
 B=-12\lambda\label{so11B2}
\end{equation}
since $\varphi^1\neq 0$. The last equation then implies $\varphi^b_C=0$. From (\ref{so11B2}) we find
\begin{equation}
 \frac{1}{||\varphi_C||^6}=24\lambda.
\end{equation}
 The value of $||\varphi_C||^2=(\varphi_C^1)^2-(\varphi_C^2)^2$ is fixed by $\lambda$ but not $\varphi_C^1$ and $\varphi_C^2$ individually. The value of the potential at these critical points is
\begin{equation}
 P^{(5)}_{TOT}\arrowvert_{\varphi^C}=\frac{3}{8||\varphi_C||^4}
\end{equation}
and therefore it corresponds to a one parameter family of deSitter ground states. The stability of the critical points is checked by calculating the eigenvalues of the Hessian of the potential, which are easily found as
\begin{equation}
 \{0,\frac{3\left[(\varphi_C^1)^2+(\varphi_C^2)^2\right]}{||\varphi_C||^8},\underbrace{\frac{1}{4||\varphi_C||^6},...,\frac{1}{4||\varphi_C||^6}}_{(\tilde{n}-2) \text{ times}}\}.\nonumber
\end{equation}
The eigenvalues are all nonnegative, thus the one parameter family of deSitter critical points is found to be stable.
\paragraph{With a universal hypermultiplet:}
Since a $U(1)_H$ hyper-gauging will not change the nature of the critical points we will just do the $SO(1,1)_H$ hyper-gauging.
\subparagraph{$SO(2)\times SU(2)_R \times SO(1,1)_H$ gauging:}
The $SO(1,1)$ gauge field is chosen as the linear combination $V_a A_\mu^a,\quad a=0,6,7,8,...,\tilde{n}$. The total potential is
\begin{equation}
 P^{(5)}_{TOT} = P^{(T)} + \lambda (P^{(R)} + P^{(H)})\nonumber
\end{equation}
where $P^{(T)}$ is as given as in the last case; $P^{(R)}$ and $P^{(H)}=2\mathcal{N}_{iA} \mathcal{N}^{iA}$ is modified with
\begin{equation}
\begin{array}{rcl}
 P^{(R)}&=&-4 C^{IJK} \vec{P}_I\cdot \vec{P}_J h_K,\\
 \mathcal{N}^{iA}&=&\frac{\sqrt{6}}{4} (h^I K_I^X +h^a V_a T^X_4) f_X^{iA}.\nonumber
\end{array}
\end{equation}
At the base point of the hyperscalar manifold, the $q$-derivatives of the total potential are found as
\begin{equation}
 \begin{array}{rcl}
  \partial_{V} P^{(5)}_{TOT}\arrowvert_{q^C}&=&0\\
\partial_{\sigma} P^{(5)}_{TOT}\arrowvert_{q^C}&=&\frac{27}{4}\lambda \varphi^5 \left\{2 V_b\varphi^b + \frac{\sqrt{2} V_0}{||\varphi||^2}\right\}\\
\partial_{\theta} P^{(5)}_{TOT}\arrowvert_{q^C}&=&\frac{27}{2}\lambda \varphi^4 \left\{2 V_b\varphi^b + \frac{\sqrt{2} V_0}{||\varphi||^2}\right\}\\
\partial_{\tau} P^{(5)}_{TOT}\arrowvert_{q^C}&=&\frac{27}{2}\lambda \varphi^3 \left\{2 V_b\varphi^b + \frac{\sqrt{2} V_0}{||\varphi||^2}\right\}\\
 \end{array}\nonumber
\end{equation}
with $b=6,...,\tilde{n}$. There are two ways of setting these expressions to zero. The first one is to set $\varphi_C^3=\varphi_C^4=\varphi_C^5=0$ and the second one is to set $2 V_b\varphi^b + \frac{\sqrt{2} V_0}{||\varphi||^2}=0$. One can show that the first case leads to the second one (and vice versa), and we choose to proceed with the second case. With this choice the $\varphi^{\tilde{x}}$-derivatives of the potential (at the base point of the hyperscalar manifold) are evaluated as
\begin{equation}
 \begin{array}{rcl}
  \partial_{\varphi^1} P^{(5)}_{TOT}\arrowvert_{\phi^C}&=&\varphi^1 \left(9\lambda - \frac{3\left[(\varphi^1)^2-(\varphi^2)^2\right]-||\varphi||^2}{4||\varphi||^8}\right),\\
\partial_{\varphi^2} P^{(5)}_{TOT}\arrowvert_{\phi^C}&=&-\varphi^2 \left(9\lambda - \frac{3\left[(\varphi^1)^2-(\varphi^2)^2\right]-||\varphi||^2}{4||\varphi||^8}\right),\\
\partial_{\varphi^d} P^{(5)}_{TOT}\arrowvert_{\phi^C}&=&\frac{3}{4}\varphi^d \left(-12\lambda+\frac{(\varphi^1)^2-(\varphi^2)^2}{||\varphi||^8}\right),\qquad d=3,...,\tilde{n}.
 \end{array}\nonumber
\end{equation}
The only way to set these equations to zero is to have $\varphi_C^d=0$ together with $\lambda=\frac{1}{18 ||\varphi||^6}$. Note that, setting $\varphi_C^d=0$ implies $V_0=0$. So, in order to have the potential term coming from the $SO(1,1)_H$ gauging not vanish we must have at least $\tilde{n}\geq 6$. Plugging in everything into the total potential, one finds that
\begin{equation}
 P^{(5)}_{TOT}\arrowvert_{\varphi^C}=\frac{3}{8||\varphi_C||^4}.\nonumber
\end{equation}
The value of $\lambda$ determines $||\varphi_C||^2=(\varphi_C^1)^2-(\varphi_C^2)^2$ but not the $\varphi_C^1$ and $\varphi_C^2$ individually. Therefore we found a one parameter family of deSitter ground states. The eigenvalues of the Hessian of the potential, evaluated at the family of critical points, are found to be
\begin{equation}\begin{array}{c}
 \{\displaystyle 0,\overbrace{\frac{1}{4||\varphi||^6},...,\frac{1}{4||\varphi||^6}}^{(\tilde{n}-3) \text{ times}},\frac{1+2 V_b V_b}{4||\varphi||^6},\\[9pt]
\displaystyle \frac{1}{8||\varphi||^4},\frac{1}{8||\varphi||^4},
\frac{1}{2||\varphi||^4},\frac{1}{2||\varphi||^4},\frac{3[(\varphi^1)^2+(\varphi^2)^2]}{||\varphi||^8}\}.
\end{array}\nonumber
\end{equation}
The eigenvalues are all nonnegative and therefore the critical points are stable.

\subsubsection{\label{so11u1rhypers}$U(1)_R$ symmetry gauging}
\paragraph{Without hypermultiplets:}
The calculation in \cite{Gunaydin:Vacua} for $\tilde{n}=3$ was later generalized to arbitrary $\tilde{n}\geq 3$ in \cite{smet}. Let us briefly quote their results. A linear combination $A_\mu [U(1)_R]=V_I A^I_\mu$ of the vector fields is taken as the $U(1)_R$ gauge field. The scalar potential is now
\begin{equation}
  P^{(5)}_{TOT}=P^{(T)}+\lambda P^{(R)}\nonumber
\end{equation}
where
\begin{equation}
 P^{(R)}=-4\sqrt{2} V_0 V_i \varphi^i ||\varphi||^{-2}+2 |V|^2 ||\varphi||^2
\end{equation}
with $i=3,...,\tilde{n}$ and $|V|^2=V_i V_i$. Demanding $\partial_{\varphi^{\tilde{x}}} P^{(5)}_{TOT}=0$, one obtains the following conditions
\begin{equation}
 \begin{array}{rcl}
  \frac{\varphi^i_C}{||\varphi_C||^4}&=&16\sqrt{2}\lambda V_0 V_i\\
\frac{1}{||\varphi_C||^6}&=&-\frac{1}{2}(16\sqrt{2}\lambda V_0 |V|)^2 +8\lambda |V|^2\label{so11u1rconds}
 \end{array}
\end{equation}
with the constraints
\begin{equation}\begin{array}{rcl}
|V|^2&>&0\\
32\lambda (V_0)^2 &<&1.\label{so11u1rconstraints}
\end{array}\end{equation}
Given a set of $V_I$ subject to (\ref{so11u1rconstraints}), we see that $||\varphi||^2$ and $\varphi^i$ (and thus $(\varphi^1)^2-(\varphi^2)^2$) are completely determined by (\ref{so11u1rconds}) but $\varphi^1$ and $\varphi^2$ are otherwise undetermined. The value of the potential at these one parameter family of critical points becomes
\begin{equation}
 P^{(5)}_{TOT}\arrowvert_{\varphi^C}=3\lambda||\varphi||^2 |V|^2 (1-32\lambda(V_0)^2)
\end{equation}
and this corresponds to deSitter vacua. The stability is checked by calculating the eigenvalues of the Hessian of the potential at the critical point. We can use the $SO(1,1)$ invariance together with the $SO(\tilde{n}-2)$ of the ${\varphi}^i$ to take for any critical point $\varphi_C = (\varphi^1,0,\varphi^3,0,...,0)$. With these choices the Hessian becomes block diagonal at the critical point. $\varphi^2$ is a zero mode and the sector $\varphi^4,...,\varphi^{\tilde{n}}$ consists of a unit matrix times $\frac{1}{4}||\varphi||^{-6}$. The only non-diagonal part of the Hessian is
\begin{equation}
 \partial_{\tilde{x}}\partial_{\tilde{y}}P^{(5)}_{TOT}\arrowvert_{\tilde{x},\tilde{y}=1,3}=\gamma \left(\begin{array}{cc}
                                                                                                                                                        (\varphi^1)^2 [6(\varphi^1)^2 + 5(\varphi^3)^2]&-\varphi^1 [8(\varphi^1)^2 \varphi^3 + 3 (\varphi^3)^3]\\
-\varphi^1 [8(\varphi^1)^2 \varphi^3 + 3 (\varphi^3)^3]&\frac{1}{4}[2(\varphi^1)^4+37 (\varphi^1)^2 (\varphi^3)^2 +5 (\varphi^3)^4]
                                                                                                                                                       \end{array}\right)\nonumber
\end{equation}
 with $\gamma = ||\varphi||^{-8} [2(\varphi^1)^2-(\varphi^3)^2]^{-1}$. The determinant and the trace of this part of the Hessian are
\begin{equation}\begin{array}{rcl}
 \text{det } \partial\partial P^{(5)}_{TOT} &=&\displaystyle  \frac{12 (\varphi^1)^6 -12 (\varphi^1)^4 (\varphi^3)^2 + 11 (\varphi^1)^2 (\varphi^3)^4}{4 ||\varphi||^{14} [2(\varphi^1)^2-(\varphi^3)^2]^2}\\[9pt]
\text{tr } \partial\partial P^{(5)}_{TOT} &=&\displaystyle  \frac{26(\varphi^1)^4 + 57 (\varphi^1)^2 (\varphi^3)^2 +5 (\varphi^3)^4}{4 ||\varphi||^8 [2(\varphi^1)^2-(\varphi^3)^2] }\nonumber
\end{array}\end{equation}
which are both positive because of $(\varphi^1)^2 > (\varphi^3)^2$ and therefore the family of critical points is found to be stable. We note that, although the above quantities are both positive, they are slightly different than the ones found in \cite{smet}, where the authors fixed the coupling constants with $\lambda=1$.

\paragraph{With a universal hypermultiplet:}
Inclusion of a hypermultiplet in the theory changes the total potential to
\begin{equation}
  P^{(5)}_{TOT}=P^{(T)} +\lambda (P^{(R)} +2\mathcal{N}_{iA} \mathcal{N}^{iA})
\end{equation}
with now
\begin{equation}
 \begin{array}{rcl}
  P^{(R)}&=&-4 C^{IJK} \vec{P}_I \cdot \vec{P}_J h_K\\
\mathcal{N}^{iA}&=&\frac{\sqrt{6}}{4} (V_I h^I Y^d T_d^X) f_X^{iA}
 \end{array}
\end{equation}
where $Y^d T_d^X$ with $d=1,2,3,8$ defines the linear combination of compact Killing vectors to be used. This potential also has a one parameter family of deSitter critical points. The $U(1)$ gauging in the hyper sector will scale $P^{(R)}$ by a positive factor at the base point of the hyperscalar manifold, which can be embedded in $V_I$'s. Because of $\frac{\partial^2 P^{(5)}_{TOT}}{\partial \varphi \partial q}\arrowvert_{q^c}=0$, as we discussed before, the only thing remains to be checked is the stability of the hyper sector. Due to the lenghtiness of the expressions, we give a particular example. Taking $Y^1=Y^2=0,Y^4=-\sqrt{3} Y^8$ we arrive at a further restriction on $V_0$ in order to maintain stability.
\begin{equation}
 64 \lambda (V_0)^2 > 1\nonumber
\end{equation}
This restriction (together with (\ref{so11u1rconstraints})) is necessary and sufficient to obtain stable dS vacua. 

\subparagraph{Example:} Using the $SO(1,1)$ and the $SO(\tilde{n}-2)$ invariances as in the previous case before we added hypers, we calculated the coordinates of the critical point for the specific case $V_0 =\frac{3}{2}, V_3 = 1, \lambda=\frac{1}{96}, Y^8 = 1$ as \begin{equation}\varphi^1 = \sqrt{7} \sqrt[6]{48},\quad\varphi^3 = \sqrt{2} \sqrt[3]{36}, \quad\nonumber\varphi^2=\varphi^4=...=\varphi^{\tilde{n}}=0.\end{equation}
 The value of the potential at this critical point is $\frac{1}{8\sqrt[3]{36}}$ and the eigenvalues of the $\frac{\partial^2 P^{(5)}_{TOT}}{ \partial q^2}$ part of the Hessian at this critical point are found to be
\begin{equation}
 \frac{4\sqrt{3}-3}{64\sqrt[3]{36}},\frac{4\sqrt{3}-3}{64\sqrt[3]{36}},\frac{2505-128\sqrt{3}}{512\sqrt[3]{36}},\frac{2505-128\sqrt{3}}{512\sqrt[3]{36}}\nonumber
\end{equation}
which are all positive. Note that it is also possible to obtain unstable critical points with different choices of $Y^d$ and $V_I$. To conclude this subsection we investigate the situation with an additional noncompact hyper-gauging.
\begin{table}[t]
\begin{minipage}{\textwidth}
\begin{tabular}{|p{2.8cm}|p{3.5cm}|p{3.5cm}|p{3.6cm}|}
\hline
& No $R$ sym. gauging&$SU(2)_R$ gauging& $U(1)_R$ gauging\\
\hline
MESGT&Minkowski\newline (supersymmetric)\newline arbitrary $\tilde{n}\geq 0$&-\newline $\tilde{n}\geq 4$&AdS\newline (supersymmetric, stable) \newline+Minkowski\newline  (nonsupersymmetric)\newline $\tilde{n}\geq 1$\\
\hline
YMESGT\newline with tensors \newline and gauge group\newline $SO(2)$&Minkowski\newline (supersymmetric)\newline $\tilde{n}\geq 3$&Minkowski\newline (supersymmetric)\newline $\tilde{n}\geq 6$&AdS\newline (supersymmetric, stable +\newline nonsupersymmetric, stable)\newline+Minkowski\newline (nonsupersymmetric)\newline $\tilde{n}\geq 3$\\
\hline
YMESGT\newline with tensors \newline and gauge group\newline $SO(1,1)$\newline (broken susy)&-\newline $\tilde{n}\geq 2$&dS\newline (stable)\newline $\tilde{n}\geq 5$&dS\newline (stable)\newline $\tilde{n}\geq 3$\\
\hline
\end{tabular}\\
\caption{Ground states of $d=5, \mathcal{N}=2$ supergravity \emph{without hypermultiplets}. The columns represent different $R$-symmetry gaugings whereas the rows represent different tensor couplings. $\tilde{n}$ denotes the minimum number of vector multiplets that must be coupled to the theory in order to make the respective gauging possible. ``-'' means there are no ground states.\label{tablenohyp}}
\end{minipage}
\end{table}

\begin{table}[tbp]
\begin{minipage}{\textwidth}
\begin{tabular}{|p{2.8cm}|p{3.5cm}|p{3.5cm}|p{3.6cm}|}
\hline& No $R$ sym. gauging& $SU(2)_R$ gauging & $U(1)_R$ gauging\\
\hline
MESGT&dS \newline (stable)&dS\newline (stable + unstable)&AdS\newline (stable$^b$ + unstable)\\
\hline
YMESGT\newline with tensors \newline and gauge group\newline $SO(2)$&dS\newline (stable\footnote{up to $n= 6$ at least})&dS\newline (stable + unstable)&AdS\newline (stable\footnote{We haven't found explicit results but the form of the expressions suggests that it is possible to obtain stable vacua.} + unstable)\\
\hline
YMESGT\newline with tensors \newline and gauge group\newline $SO(1,1)$&-&dS\newline (stable)&dS\newline (stable + unstable)\\
\hline
\end{tabular}
\caption{Ground states of $d=5, \mathcal{N}=2$ supergravity \emph{with one hypermultiplet} and \emph{with noncompact $SO(1,1)$ gauging of the hyper sector}. The columns represent different $R$-symmetry gaugings whereas the rows represent different tensor couplings. Note that noncompact hyper-gauging implies broken supersymmetry. The Minkowskian ground states are not listed. ``-'' means there are no ground states.\label{tablewithhyp}}
\end{minipage}
\end{table}

\subparagraph{$SO(1,1)\times U(1)_R \times SO(1,1)_H$ gauging:}
This is very similar to the previous case. The only difference is
\begin{equation}
 \mathcal{N}^{iA}=\frac{\sqrt{6}}{4}(V_I h^I Y^d T_d^X+W_I h^I T_4^X) f_X^{iA}.
\end{equation}
The linear combination $A_\mu[SO(1,1)]=W_I A^I_\mu$ of the vector fields is used as the $SO(1,1)$ gauge field. The $SO(1,1)$ coupling constant is absorbed in $W_I$'s and the fact that the $U(1)_R$ gauge vector field must be orthogonal to the $SO(1,1)_H$ gauge field tells the orthogonality condition
\begin{equation}
 V_I W_I = 0.
\end{equation}
The first derivatives of the potential can be set to zero by using (\ref{so2d1}), (\ref{so2fracv}) and $W_0 = -\sqrt{2} W_b \varphi^b ||\varphi||^2,\quad b = 3,...,\tilde{n}$. We found that the potential has a one parameter family of deSitter ground states. The stability again depends on the values taken for the constants. The calculation is quite messy and here we will look at a particular example with stable ground state.

\subparagraph{Example:} For the constants in the theory, we take the following numbers:$ V_0 = \frac{1}{2\sqrt{2}};$ $V_3 =\frac{1}{4};$  $ \text{\mbox{$V_e=W_e=0, (e = 4,...,\tilde{n})$}}; \lambda = 2+\sqrt{3}; W_0 = \frac{1}{96\sqrt{2(2+\sqrt{3})}}; W_3 = -\frac{(2+\sqrt{3})^{3/2}}{96(7+4\sqrt{3})};$ $ Y^1=Y^2=0; Y^5=-\frac{Y^4}{\sqrt{3}}=\frac{1}{3}\sqrt{\frac{5}{2}(2-\sqrt{3})}$. The point $\varphi^1 = 2, \varphi^2=1,\varphi^3=1,\varphi^e=0$ is a critical point. The value of the potential at this critical point is $\frac{7}{128}$ and the eigenvalues of the Hessian at this critical point become $\frac{481}{128},\frac{481}{128},\frac{1}{576}\left(490\pm \sqrt{216430}\right),0,...,0$, which are all nonnegative and hence this corresponds to a stable deSitter vacuum.

\subsection{Summary}
Our results are summarized in \emph{Table \ref{tablenohyp}} for the theories that do not include hypers and \emph{Table \ref{tablewithhyp}} for theories that include a universal hypermultiplet and have noncompact $SO(1,1)_H$ gauging of hyper isometries. 

\section{Magical Jordan Family}
It is possible to apply the results obtained for the generic Jordan family to the Magical Jordan family, provided that there are enough vector fields in the Magical family member to do the gauging that is being ``imported'' from the generic family. For instance, the $SO(2)\times SU(2)_R$ gauging requires at least $\tilde{n}=6$ vector multiplets and thus this generic family model cannot be embedded in the smallest member of the Magical Jordan family $\mathcal{M}=\frac{SL(3,\mathbb{R})}{SO(3)}$. It can only be embedded into the bigger members of the family. In this section we will see two examples of such embeddings. These models also contain other critical points, that are special to the magical family case, i.e. that were not obtained in the generic case. 

What is more interesting for the Magical Jordan family is, a non-Abelian gauging of the isometry group will introduce a potential term due to the tensor coupling. This is because of the fact that, unlike in the generic Jordan family theories, in the Magical Jordan family theories there are vector fields that are nontrivially charged under the non-Abelian gauge group. By ``nontrivial'', we mean that there are other vectors, than the ones in the adjoint representation of the gauge group $K$, that are not singlets. These vector fields should be dualized to tensor fields and this dualization introduces a scalar potential $P^{(T)}$. An example to such a gauging will be investigated in the last part of this section.

\subsection{$\mathcal{M}=SL(3,\mathbb{R})/SO(3)$}
$\mathcal{M}$ is described by the hypersurface $N(h)=C_{\tilde{I}\tilde{J}\tilde{K}}h^{\tilde{I}}h^{\tilde{J}}h^{\tilde{K}}=1$ of the cubic polynomial
\begin{equation}
N(h)=\frac{3}{2}\sqrt{3}h^3\eta_{IJ}h^{I}h^J+\frac{3\sqrt{3}}{2\sqrt{2}}\gamma_{IMN}h^I h^M h^N,
\end{equation}
where
\begin{eqnarray}
\begin{array}{rclrcl}
I,J&=&0,1,2&M,N&=&4,5\\
\eta_{IJ}&=&diag(+,-,-)&\gamma_0&=&-1_2\\
\gamma_1&=&\sigma_1&\gamma_2&=&\sigma_3
\end{array}\nonumber
\end{eqnarray}
In this parametrization the non-vanishing $C_{\tilde{I}\tilde{J}\tilde{K}}$'s are
\begin{eqnarray}
C_{003}=-C_{113}=-C_{223}&=&\frac{\sqrt{3}}{2}\nonumber\\
C_{044}=C_{055}&=&-\frac{\sqrt{3}}{2\sqrt{2}}\nonumber\\
C_{244}=-C_{255}&=&\frac{\sqrt{3}}{2\sqrt{2}}\nonumber\\
C_{145}&=&\frac{\sqrt{3}}{2\sqrt{2}}\nonumber
\end{eqnarray}
and their permutations. $N(h)$ indeed is the determinant of the Jordan algebra $J_3^{\mathbb{R}}$ element 
\begin{equation}
\mathbf{\tilde{h}}=\frac{\sqrt{3}}{\sqrt{2}}\left( \begin{array}{ccc}
                                 h^0+h^2&h^1&h^4\\
				 h^1&h^0-h^2&h^5\\
				 h^4&h^5&\sqrt{2}h^3\end{array}\right).\nonumber
\end{equation}
To solve $N(h)=1$ we take the parametrization:
\begin{equation}
h^I=\sqrt{\frac{2}{3}}x^I,\quad
h^M=\sqrt{\frac{2}{3}}b^M,\quad
h^3=\frac{1-b^T\bar{x}b}{\sqrt{3}||x||^2}.\nonumber
\end{equation}
where $||x||^2=\eta_{IJ}x^I x^J$ and $b^T\bar{x}b=b^M x^I \gamma_{IMN} b^N$.
\subsubsection{$SO(2)\times U(1)_R$ gauging} 
The fields $h^0$ and $h^3$ are chargeless under the action of the compact $SO(2)$ generator
\begin{equation}
\tilde{\sigma}_2 =\left( \begin{array}{ccc}
                        0&1&0\\-1&0&0\\0&0&0\end{array}\right)\nonumber
\end{equation}
while $h^1$ and $h^2$ are forming a doublet with charge $2$; and $h^4$ and $h^5$ are forming another doublet with charge $1$. To gauge the $SO(2)$ subgroup of the symmetry group we will use a linear combination of the fields $A^0_\mu$ and $A^3_\mu$. The $U(1)_R$ symmetry will be gauged by another linear combination of the same fields. We will split the indices as $i,j=0,3$; $m,n=1,2,4,5$.

The symplectic matrix $\Omega_{mn}$ that appears in the potential is given by
\begin{equation}
\Omega_{mn}=\left(\begin{array}{rrrr}
                    0&1&0&0\\-1&0&0&0\\
                    0&0&0&1\\0&0&-1&0\end{array}\right)\nonumber
\end{equation}
and the non-vanishing $\Lambda_i^{mn}$'s are
\begin{equation}
\Lambda_0^{44}=\Lambda_0^{55}=-\frac{1}{2},\quad
\Lambda_3^{11}=\Lambda_3^{22}=\frac{1}{\sqrt{2}}.\nonumber
\end{equation}

Now the potential terms $P^{(T)} = \frac{3\sqrt{6}}{16}h^i\Lambda_i^{mn}h_m h_n$ and $P^{(R)} = -4 C^{ijk} V_i V_j h_k$ become:
\begin{equation}
P^{(T)}=\frac{3\sqrt{6}}{16}[-h^0 [(h_4)^2 + (h_5)^2] + \sqrt{2}{h^3} [(h_1)^2 + (h_2)^2)]\arrowvert_{N(h)=1}\label{magicptso2}
\end{equation}
\begin{equation}
P^{(R)}=-2V_0\left\{V_0 ||x||^2 + \sqrt{2} V_3 \left(2 x^0 \frac{1-b^T\bar{x}b}{||x||^2} - [(b^4)^2+(b^5)^2]\right)\right\}\label{magicprso2}
\end{equation}
Here we defined $h_{\tilde{I}} \equiv \frac{1}{3} \frac{\partial}{\partial h^{\tilde{I}}} N \arrowvert_{N=1}$. One should note that determinants of the vector/tensor field metric $\stackrel{o}{a}_{\tilde{I}\tilde{J}}$ and the hypersurface metric $g_{\tilde{x}\tilde{y}}$ are given by $1$ and $\frac{243}{16||x||^4}$, respectively. This tells us that both metrics are positive definite on the scalar manifold when $||x||^2\neq 0$.

The total potential is:
\begin{equation}
P^{(5)}=P^{(T)}+\lambda P^{(R)}
\end{equation}
where $P^{(T)}$ and $P^{(R)}$ were given in (\ref{magicptso2}),(\ref{magicprso2}). It is quite difficult to calculate the most general solution for the critical points of this potential. Instead, we look at specific sectors.\\[5pt]
{\bf\underline{Sector 1: $b^4=b^5=0$ at the critical point}}

This sector looks quite similar to the generic case with $\frac{SO(2,1)\times SO(1,1)}{SO(2)}$ as the scalar manifold with $SO(2)\times U(1)_R$ gauging, yet there is an important difference. In the generic case, to have the metrics $\stackrel{o}{a}_{\tilde{I}\tilde{J}}$ and $g_{\tilde{x}\tilde{y}}$ positive definite, we were forced to look at the sector $h^0\ne 0$. Now this restriction does not apply anymore and this will help us find more ground states.

To find the critical points of the scalar manifold, we take the derivatives of the total potential with respect to all the scalars of the manifold (i.e. $x^0, x^1, x^2, b^4, b^5$) and set them equal to zero. These derivatives are given by
\begin{eqnarray}\begin{array}{rcl}
P^{(5)*}_{,0}&=& -(A+\lambda B) x^0 +\frac{x^0}{4||x||^6}-4\sqrt{2}\lambda \frac{V_0 V_3}{||x||^2}\\
P^{(5)*}_{,1}&=& (A+\lambda B) x^1\\
P^{(5)*}_{,2}&=& (A+\lambda B) x^2\\
P^{(5)*}_{,4}&=& P^{(5)}_{,5}=0\label{so2b4b50}
\end{array}
\end{eqnarray}
where
\begin{eqnarray}\begin{array}{rcl}
A&=&\displaystyle \frac{3}{4}\frac{(x^1)^2+(x^2)^2}{||x||^8}+\frac{1}{4||x||^6}\\[8pt]
B&=&\displaystyle 4V_0 (V_0-\frac{2\sqrt{2}V_3 x^0}{||x||^4})\nonumber
\end{array}\end{eqnarray}
and $*:\text{at } b^4=b^5=0$. The case with $x^0 \ne 0$ is almost equivalent to the generic case and it was studied in section \ref{genericso2u1r}. There are two possibilities for setting the first derivatives of the potential to zero.

\underline{Case 1: $x^1_C=x^2_C=0$ }
Letting $V_0=0$, we can get a Minkowski ground state with broken supersymmetry. Or one can set the above equations to zero by letting $\sqrt{2} V_3 = V_0 (x^0)^3$. The value of the potential then becomes $P^{(5)}\arrowvert_{\varphi_C}=-6\lambda (V_0 x^0)^2$ and this corresponds to a stable, supersymmetric Anti-deSitter critical point. The same analysis we did in section \ref{genericso2u1r} shows that this critical point is a saddle point.

\underline{Case 2: $(x^1_C)^2+( x^2_C)^2\neq 0$ }
Setting the first derivatives of the potential to zero, one arrives at (c.f. eqn's (\ref{so2varphi1},\ref{so2arrive}))
\begin{equation}
 \begin{array}{rcl}
 \displaystyle  \frac{x^0}{||x||^4}&=&16\sqrt{2}\lambda V_0 V_3\\[8pt]
\displaystyle \frac{1}{||x||^6}&=&\displaystyle \frac{1}{2}(16\sqrt{2}\lambda V_0 V_3)^2 +8\lambda (V_0)^2.
 \end{array}
\end{equation}
In order these equations to be consistent, one needs
\begin{equation}
 32\lambda (V_3)^2 >1,\qquad V_0 V_3 \neq 0.
\end{equation}
There is a one parameter family of Anti-deSitter ground states which do not preserve the full $\mathcal{N}=2$ supersymmetry, and the value of the potential at these critical points is given by
\begin{equation}
 P^{(5)}\arrowvert_{\varphi_C}=-\frac{3}{8}\frac{1}{||x||^4}<0.
\end{equation}

Now we come to the case with $x^0=0$. Equations (\ref{so2b4b50}) reduce to
\begin{equation}
 \begin{array}{rcl}
V_0 V_3 &=& 0,\\
\displaystyle \frac{1}{2}x^a\left(8\lambda (V_0)^2 + \frac{1}{||x||^6}\right)&=&0,\qquad a=1,2.
 \end{array}
\end{equation}
It's not possible to solve the second equation by letting $V_0$, so one sets $V_3=0$. The second equation then is solved by $8\lambda (V_0)^2 ||x||^6= - 1$. The potential evaluated at the critical point is given by $P^{(5)}\arrowvert_{\varphi_C}=-\frac{3}{8}\frac{1}{[(x^1)^2+(x^2)^2]}$ and this is a one parameter family of stable and AdS vacua with broken supersymmetry ($x^a_C \neq 0\Rightarrow P^a \neq 0$). \\[5pt]
{\bf\underline{Sector 2: $x^1=x^2=0$ at the critical point}}

The derivative of the potential with respect to the scalar $x^2$ is
\begin{equation}
 P^{(5)}_{,2}=-\frac{[(b^4)^2-(b^5)^2] [4+8 D +5 D^2-128\sqrt{2}\lambda V_0 V_3 (x^0)^3 - 8 (x^0)^6]}{32(x^0)^4}
\end{equation}
where $D=[(b^4)^2+(b^5)^2] x^0$.\\[3pt]
\underline{Case 1: $(b^4)^2-(b^5)^2=0$ }\\[2pt]
Setting $b^4=b^5$, the $b^4$ and $b^5$ derivatives of the potential become
\begin{equation}
  P^{(5)}_{,4}= P^{(5)}_{,5}=\frac{b^5 \left\{(b^5)^2 + 3 (b^5)^4 x^0 - (x^0)^2 (16\sqrt{2}\lambda V_0 V_3+(x^0)^3)\right\}}{4 (x^0)^2}
\end{equation}
One way to set this equal to zero is to make $b^5=0$. But this means $b^4=0$ and this case was already covered in Sector 1. Instead, we set
\begin{equation}
  (b^5)^2 + 3 (b^5)^4 x^0 = (x^0)^2 (16\sqrt{2}\lambda V_0 V_3+(x^0)^3) 
\end{equation}
Plugging this in the $x^1$ derivative of the potential yields
\begin{equation}
 P^{(5)}_{,1}=\frac{(b^5)^2 \left\{-1+(b^5)^2 x^0 (-2+(b^5)^2 x^0)\right\}}{4 (x^0)^2}.
\end{equation}
This vanishes if we set $x^0=\frac{1\pm\sqrt{2}}{(b^5)^2}$. Plugging this into the $x^0$ derivative of the potential, one finds
\begin{equation}
 P^{(5)}_{,0}=-\frac{10 \pm 7\sqrt{2} + 16 (1\pm\sqrt{2})\lambda (V_0)^2)}{4 (b^5)^2}.
\end{equation}
This can only vanish if one selects the lower sign. One finds the value of the ratio $\lambda$ of the coupling constants as a function of $V_0$ as
\begin{equation}
 \lambda =\frac{10-7\sqrt{2}}{16(\sqrt{2}-1) (V_0)^2}.\label{magicso2lambda}
\end{equation}
The value of the potential at the critical point is given by
\begin{equation}
 P^{(5)}\arrowvert_{\varphi_C}=\frac{3 \left(10-7\sqrt{2}-(b^5)^{12}\right)}{8 (b^5)^4}
\end{equation}
and $b^5$ is determined as a solution to the equation
\begin{equation}
 2 V_3 (b^5)^6=\left(\sqrt{2}-1-(4+3\sqrt{2}) (b^5)^{12}\right) V_0
\end{equation}
So, the sign of the value of the potential at the critical point can be tuned by carefully choosing $V_0$ and $V_3$. Using Mathematica, we found that the Hessian has at least one positive and one negative eigenvalue for any choice of $V_0$ and $V_3$ and hence the critical point is a saddle point. The same result can be obtained by setting $b^4=-b^5$. \\[3pt]
\underline{Case 2: $4+8 D +5 D^2=128\sqrt{2}\lambda V_0 V_3 (x^0)^3 + 8 (x^0)^6$ }\\[2pt]
In this case the $x^1$-derivative of the potential vanishes and the $b^4$ and $b^5$-derivatives reduce to
\begin{equation}
\frac{P^{(5)}_{,4}}{b^4}= \frac{P^{(5)}_{,5}}{b^5}=\frac{-4+[(b^4)^2+(b^5)^2] x^0 \{-4 +[(b^4)^2+(b^5)^2] x^0\}}{32 (x^0)^3}.
\end{equation}
Setting $x^0=\frac{2\pm 2\sqrt{2}}{(b^4)^2+(b^5)^2}$ makes this expression vanish. Plugging this into the $x^0$-derivative equation yields
\begin{equation}
P^{(5)}_{,0}=-\frac{10 \pm 7\sqrt{2} + 16 (1\pm\sqrt{2})\lambda (V_0)^2)}{2 [(b^4)^2+(b^5)^2]}.
\end{equation}
Again, as in the last case, this can only vanish by selecting the lower sign. This sets the value of $\lambda$ as given in (\ref{magicso2lambda}). The value of the potential at the critical point is found as
 \begin{equation}
  P^{(5)}\arrowvert_{\varphi_C}=\frac{3\left(640-448\sqrt{2}-[(b^4)^2+(b^5)^2]^6\right)}{128[(b^4)^2+(b^5)^2]^2}.
 \end{equation}
$(b^4)^2+(b^5)^2$ can be tuned by carefully choosing $V_0$ and $V_3$ as in the last case, but $b^4$ and $b^5$ are otherwise not determined. Using Mathematica, we found that the Hessian of the potential at this one parameter family of critical points has both positive and negative eigenvalues for any choice of $V_0$ and $V_3$, as in the last case; so the ground states are ``saddle curves'',i.e. they are neither minima or maxima.

\subsubsection{$SO(1,1)\times U(1)_R$ gauging} 
The fields $h^1$ and $h^3$ are chargeless under the action of the non-compact $\frac{SL(2,\mathbb{R})}{SO(2)}$ generator
\begin{equation}
\tilde{\sigma}_3 =\left( \begin{array}{ccc}
                        1&0&0\\0&-1&0\\0&0&0\end{array}\right)\nonumber
\end{equation}
while $h^0$ and $h^2$ are forming a doublet with charge $2$; and $h^4$ and $h^5$ are forming another doublet with charge $1$. To gauge the $SO(1,1)$ subgroup of the symmetry group we will use a linear combination of the fields $A^1_\mu$ and $A^3_\mu$. The $U(1)_R$ symmetry will be gauged by another linear combination of the same fields. We will split the indices as $i,j=1,3$; $m,n=0,2,4,5$.

The symplectic matrix $\Omega_{mn}$ that appears in the potential is given by
\begin{equation}
\Omega_{mn}=\left(\begin{array}{rrrr}
                    0&1&0&0\\-1&0&0&0\\
                    0&0&0&1\\0&0&-1&0\end{array}\right)\nonumber
\end{equation}
and the non-vanishing $\Lambda_i^{mn}$'s are
\begin{equation}
\Lambda_3^{00}=\frac{1}{\sqrt{2}},\quad
\Lambda_3^{22}=-\frac{1}{\sqrt{2}},\quad
\Lambda_1^{45}=\frac{1}{2},\quad
\Lambda_2^{54}=\frac{1}{2}.\nonumber
\end{equation}

Now the potential terms $P^{(T)} = \frac{3\sqrt{6}}{16}h^i\Lambda_i^{mn}h_m h_n$ and $P^{(R)} = -4 C^{ijk} V_i V_j h_k$ become:
\begin{equation}
P^{(T)}=\frac{3\sqrt{6}}{16}[h^1 h_4 h_5 + \frac{h^3}{\sqrt{2}} ((h_0)^2 - (h_2)^2)]\arrowvert_{N(h)=1}\label{ptso11}
\end{equation}
\begin{equation}
P^{(R)}=2V_1[V_1 ||x||^2 + 2V_3(-\sqrt{2}x^1 \frac{1-b^T\bar{x}b}{||x||^2}+\sqrt{2}b^4 b^5)]\label{prso11}
\end{equation}
Here we defined $h_{\tilde{I}} \equiv \frac{1}{3} \frac{\partial}{\partial h^{\tilde{I}}} N \arrowvert_{N=1}$. One should note that determinants of the vector/tensor field metric $\stackrel{o}{a}_{\tilde{I}\tilde{J}}$ and the hypersurface metric $g_{\tilde{x}\tilde{y}}$ are given by $1$ and $\frac{243}{16||x||^4}$, respectively. This tells us that both metrics are positive definite on the scalar manifold when $||x||^2\neq 0$.

The total potential is:
\begin{equation}
P^{(5)}=P^{(T)}+\lambda P^{(R)}
\end{equation}
where $P^{(T)}$ and $P^{(R)}$ were given in (\ref{ptso11}),(\ref{prso11}). It is quite difficult to calculate the most general solution for the critical points of this potential. Instead, we look at specific sectors.\\

{\bf\underline{Sector 1: $b^4=b^5=0$ at the critical point}}

This sector looks quite similar to the generic case with $\frac{SO(2,1)\times SO(1,1)}{SO(2)}$ as the scalar manifold with $SO(1,1)\times U(1)_R$ gauging, yet there is an important difference. In the generic case, to have the metrics $\stackrel{o}{a}_{\tilde{I}\tilde{J}}$ and $g_{\tilde{x}\tilde{y}}$ positive definite, we were forced to look at the sector $h^0\ne 0$. Now this restriction does not apply anymore and this will help us find more ground states. 

To find the critical points of the scalar manifold, we take the derivatives of the total potential with respect to all the scalars of the manifold (i.e. $x^0, x^1, x^2, b^4, b^5$) and set them equal to zero. These derivatives are given by
\begin{eqnarray}\begin{array}{rcl}
P^{(5)*}_{,0}&=& (A+\lambda B) x^0\\
P^{(5)*}_{,1}&=& -(A+\lambda B) x^1+\frac{x^1}{4||x||^6}-4\sqrt{2}\lambda \frac{V_1 V_3}{||x||^2}\\
P^{(5)*}_{,2}&=& -(A+\lambda B) x^2\\
P^{(5)*}_{,4}&=& P^{(5)}_{,5}=0\label{b4b50}
\end{array}
\end{eqnarray}
where
\begin{eqnarray}\begin{array}{rcl}
A&=&\displaystyle \frac{||x||^2-3[(x^0)^2-(x^2)^2]}{4||x||^8}\\[9pt]
B&=&\displaystyle 4V_1 (V_1+\frac{2\sqrt{2}V_3 x^1}{||x||^4})\nonumber
\end{array}\end{eqnarray}
and $*:\text{at } b^4=b^5=0$. The case with $x^0 \ne 0$ is almost equivalent to the generic case and it was studied in section \ref{so11u1rhypers}. Setting the first derivatives of the potential to zero, one arrives at (c.f. eqn (\ref{so11u1rconds}))
\begin{equation}\begin{array}{rcl}
\displaystyle  \frac{1}{||x||^6}&=&-256 (\lambda V_1 V_3)^2 + 8 \lambda (V_1)^2,\\[9pt]
\displaystyle \frac{x^1}{||x||^4}&=&16\sqrt{2}\lambda V_1 V_3.\label{magicso11u1rconds}
\end{array}\end{equation}
There is a one parameter family of ground states where the potential at the critical point is given by
\begin{equation}
P^{(5)}\arrowvert_{\varphi_C}=3\lambda ||x||^2 (V_1)^2 (1-32\lambda (V_3)^2),
\end{equation} 
with $V_1\ne 0$. Choosing $1>32\lambda (V_3)^2$ leads to $||x||^2>0$, whereas $1<32\lambda (V_3)^2$ leads to $||x||^2<0$, therefore this family of ground states correspond to deSitter vacua.

It was found in the generic case that this family of deSitter vacuum is stable. Let us check if the stability applies to the magical model at hand. The stability is manifested by the positivity of the eigenvalues of the Hessian of the potential. The Hessian, evaluated at the family of critical points, is of the block diagonal form
\begin{equation}
(\partial\partial P^{(5)})\arrowvert_{\varphi_C} =\left[\begin{array}{c|c}
E_{3\times 3}&0\\
\hline
0&F_{2\times 2}
\end{array}\right]\label{hesEF}
\end{equation}

$E$ has one zero eigenvalue. The product of the remaining two eigenvalues is $\frac{3((x^0)^2+(x^2)^2)}{4||x||^{14}}$. This tells us one must have $||x||^2>0$ in order to have positive eigenvalues. Note the fact that in the generic case this restriction came from the positivity rule of the metrics $\stackrel{o}{a}_{\tilde{I}\tilde{J}}$ and $g_{\tilde{x}\tilde{y}}$ whereas now it is a requirement to have the eigenvalues of the Hessian positive definite. The sum of the eigenvalues of $E$ is
\begin{equation}
Tr(E)=\frac{13(x^0)^4-3(x^1)^4-14(x^1)^2(x^2)^2-11(x^2)^2+(x^0)^2(22(x^1)^2-2(x^2)^2)}{4||x||^{10}}.\nonumber
\end{equation}
This is positive definite in the region $||x||^2>0$.

The product of the eigenvalues of $F$ is
\begin{equation}
Det(F)=\frac{1}{16}[-(x^1)^2||x||^4+\frac{1}{||x||^6}(\frac{(x^1)^2}{||x||^2}+2)^2-\frac{2(x^1)^2}{||x||^2}(\frac{(x^1)^2}{||x||^2}+2)]\nonumber
\end{equation}
and their sum is
\begin{equation}
Tr(F)=\frac{1}{2}x^0[\frac{2}{||x||^4}+(x^1)^2(\frac{1}{||x||^6}-2)].
\end{equation}

These two quantities are not positive definite everywhere on the domain of the family of deSitter vacua. In the region where $x^1$ and $x^2$ are close to zero their limits are
\[
\lim_{x^1,x^2\rightarrow 0}Det(F)=\frac{1}{4(x^0)^6}\qquad \lim_{x^1,x^2\rightarrow 0}Tr(F)=\frac{1}{(x^0)^3}.
\]
These are positive in the region $x^0>0$. In the region where $||x||^2$ is close to zero the limits become
\[
\lim_{||x||^2\rightarrow 0}Det(F)\rightarrow\frac{1}{16||x||^6}(\frac{(x^1)^2}{||x||^2}+2)^2\qquad \lim_{||x||^2\rightarrow 0}Tr(F)\rightarrow\frac{x^0(x^1)^2}{2||x||^6}.
\]
Again, these are positive in the region $x^0>0$. Thus these two regions on the scalar manifold contain stable deSitter vacua. There is also the relations \ref{magicso11u1rconds} that tell us $x^1$ and $[(x^0)^2-(x^2)^2]$ can be tuned by a careful choice of $V_i$'s. But $x^0$ and $x^2$ are otherwise not fixed.Although $x^0$ is not fixed, it is not possible to make a transitions between $x^0>0$ and $x^0<0$ and at the same time to keep $||x||^2>0$ fixed. Hence, in reality only the second case, where $||x||^2$ is small and positive, and $x^0>0$, is a stable deSitter ground state.

We now look at the case where $x^0=0$ at the critical point. To make the third expression in (\ref{b4b50}) zero there are two possibilities:

\underline{1) $A+\lambda B=0$:}
The second expression gives us:
\begin{equation}
\frac{x^1}{||x||^4}=16\sqrt{2}\lambda V_1 V_3.\label{x1b4b50}
\end{equation}
Plugging this back into $A+\lambda B=0$ we find
\begin{equation}
\frac{1}{||x||^6}=-256(\lambda V_1 V_3)^2 +8 \lambda (V_1)^2
\end{equation}
The left hand side of this equation is negative definite. Hence $V_1 \ne 0$ and also $32\lambda (V_3)^2>1$ and by (\ref{x1b4b50}), $x^1\ne 0$. The potential evaluated at the critical point is given by
\begin{equation}
P^{(5)}\arrowvert_{\varphi_C}=3\lambda (V_1)^2 (1-32\lambda (V_3)^2) ||x||^2
\end{equation}
which is positive, hence we have another deSitter ground state, but this is unstable as explained before, for it leads to the same Hessian (\ref{hesEF}).

\underline{2) $x^2_C=0$}: In this case $||x||^2=-(x^1)^2$. From the second expression in (\ref{b4b50}) we find
\begin{equation}V_1 = 0 \qquad\text{or}\qquad (x^1)^3 =-\frac{\sqrt{2}V_3}{V_1}\label{negdee}\end{equation}
at the critical point. Setting $V_1=0$ leads to a one parameter family of Minkowski ground states with broken supersymmetry (unless $V_3=0$, which turns off the $U(1)_R$ potential). For the other case the potential at this point becomes
\begin{equation}P^{(5)}\arrowvert_{\varphi_C}=\frac{6\sqrt{2}\lambda V_1 V_3}{x^1}\end{equation}
which is negative definite because of (\ref{negdee}) and hence this corresponds to a supersymmetric Anti-deSitter critical point. The Hessian of the scalar potential evaluated at the critical point is given by
\begin{equation}
 \text{diag}(-\frac{1+32\lambda (V_3)^2}{4 (x^1)^6},-\frac{24\lambda (V_3)^2}{(x^1)^6},\frac{1+32\lambda (V_3)^2}{4 (x^1)^6}, \frac{8\lambda (V_3)^2 }{(x^1)^3} + \frac{(x^1)^3}{4},\frac{8\lambda (V_3)^2 }{(x^1)^3} + \frac{(x^1)^3}{4} ),\nonumber
\end{equation}
which has both positive and negative eigenvalues and therefore the critical point is a saddle point of the potential.\\

{\bf\underline{Sector 2: $x^1=x^2=0$ at the critical point}}

The derivative of the potential with respect to the scalar $x^2$ is
\begin{equation}
P^{(5)}_{,2}\arrowvert_{\varphi^C}=-\frac{[(b^4)^2-(b^5)^2][2+2x^0 [(b^4)^2+(b^5)^2]+ (x^0)^2 (b^4)^2 (b^5)^2]}{8(x^0)^2}
\end{equation}
There are 4 possible ways of setting this equal to zero, i.e.
\begin{equation}
b^4=\pm b^5,\qquad b^4=\pm\frac{\sqrt{-2-2(b^5)^2 x^0}}{\sqrt{2x^0+(b^5)^2 (x^0)^2}}
\end{equation}

Inserting either of the last two values for $b^4$ into the $x^0$-derivative of the potential, one finds
\begin{equation}
P^{(5)}_{,0}\arrowvert_{\varphi^C}=4\lambda (V_1)^2 x^0
\end{equation}
But since $x^0\ne 0$, because otherwise the potential diverges at this point, one must have $V_1=0$, i.e. the $U(1)_R$ potential must be turned off. Applying this to the $b^5$-derivative of the potential, one finds
\begin{equation}
P^{(5)}_{,5}\arrowvert_{\varphi^C}=-\frac{(b^5)^3 (1+(b^5)^2 x^0)^2}{2 (x^0)^2 (2+(b^5)^2 x^0)^2}
\end{equation}
Setting this equal to zero requires either $b^5=0$ or $1+(b^5)^2 x^0=0$ but both options make the potential vanish at the critical point, so these critical points are Minkowskian.\\[5pt]

For the $b^4=b^5$ case we first note that $P^{(5)}_{,2}\arrowvert_{\varphi_C}=0$. Furthermore
\begin{equation}
P^{(5)}_{,4}\arrowvert_{\varphi^C}=P^{(5)}_{,5}\arrowvert_{\varphi^C}=\frac{b^5[2+5(b^5)^2 x^0 +3 (b^5)^4 (x^0)^2 + 16\sqrt{2}\lambda V_1 V_3 (x^0)^3]}{4(x^0)^3}\label{p4p5_x1x20}
\end{equation}
One way to set this equal to zero is to make $b^5=0$. But this means $b^4=0$ and this case was already covered in Sector 1. In particular, the potential at the critical point in this case is (note that the restrictions $V_1\ne 0,\quad 1>32\lambda (V_3)^2$ apply),
\begin{equation}
P^{(5)}\arrowvert_{\varphi_C}=3\lambda (V_1)^2 (x^0)^2
\end{equation}
which is nonnegative.

Another way to set (\ref{p4p5_x1x20}) equal to zero is to have
\begin{equation}
2+5(b^5)^2 x^0 +3 (b^5)^4 (x^0)^2 = -16\sqrt{2}\lambda V_1 V_3 (x^0)^3\label{p4p5xxxx}
\end{equation}

Plugging this in $P^{(5)}_{,1}=0$ and solving the resulting expression together with $P^{(5)}_{,0}=0$ we find
\begin{equation}
x^0=-\frac{a}{(b^5)^2}\label{p4p5yyyy}
\end{equation}
and
\begin{equation}
(x^0)^6=\frac{2-6a+5a^2-a^3}{a}\label{p4p5zzzz}
\end{equation}
where $a=16\lambda (V_1)^2$.
Plugging (\ref{p4p5xxxx}), (\ref{p4p5yyyy}) and (\ref{p4p5zzzz}) into the potential we get
\begin{equation}
P^{(5)}\arrowvert_{\varphi_C}=\frac{3-6a+3a^3}{8(x^0)^4}\label{p4p5potevall}
\end{equation}

Since the left hand side of (\ref{p4p5zzzz}) is positive definite $a$ is constrained as:
\[ 0 < a < 2-\sqrt{2}\qquad \text{or}\qquad 1 < a < 2+\sqrt{2} 
\]
In both regions the value of the potential at the critical point (\ref{p4p5potevall}) is positive, thus the critical point corresponds to a deSitter vacuum. Unfortunately, using Mathematica we found that none of these choices for $a$ yields non-negative eigenvalues for the Hessian of the potential (some eigenvalues are always negative); therefore this deSitter vacuum is unstable.

For the $b^4=-b^5$ case the value of the potential at the critical point is the same and the critical point is unstable as well.\\[12pt]

\subsection{$\mathcal{M}=SL(3,\mathbb{C})/SU(3)$}
$\mathcal{M}$ is described by the hypersurface $N(h)=C_{\tilde{I}\tilde{J}\tilde{K}}h^{\tilde{I}}h^{\tilde{J}}h^{\tilde{K}}=1$ of the cubic polynomial
\begin{equation}
N(h)=\frac{3}{2}\sqrt{3}h^4\eta_{IJ}h^{I}h^J+\frac{3\sqrt{3}}{2\sqrt{2}}\gamma_{IMN}h^I h^M h^N,
\end{equation}
where
\begin{eqnarray}
\begin{array}{rclrcl}
I,J&=&0,1,2,3&M,N&=&5,6,7,8\\
\eta_{IJ}&=&diag(+,-,-,-)\\
\gamma_0&=&-1_4&\gamma_1&=&1_2\otimes\sigma_1\\
\gamma_2&=&\sigma_2\otimes\sigma_2&\gamma_3&=&1_2\otimes\sigma_3
\end{array}
\end{eqnarray}
The non-vanishing $C_{IJK}$'s are
\begin{eqnarray}
C_{004}=-C_{114}=-C_{224}=-C_{334}&=&\frac{\sqrt{3}}{2}\nonumber\\
C_{055}=C_{066}=C_{077}=C_{088}&=&-\frac{\sqrt{3}}{2\sqrt{2}}\nonumber\\
C_{355}=C_{377}=-C_{366}=-C_{388}&=&\frac{\sqrt{3}}{2\sqrt{2}}\nonumber\\
C_{156}=C_{178}=-C_{258}=C_{267}&=&\frac{\sqrt{3}}{2\sqrt{2}}
\end{eqnarray}
and their permutations. $N(h)$ is the determinant of the Jordan algebra $J_3^{\mathbb{C}}$ element 
\begin{equation}
\mathbf{\tilde{h}}=\frac{\sqrt{3}}{\sqrt{2}}\left( \begin{array}{ccc}
                                 h^0-h^3&h^1+ih^2&h^6-ih^8\\
				 h^1-ih^2&h^0+h^3&h^5-ih^7\\
				 h^6+ih^8&h^5+ih^7&\sqrt{2}h^4\end{array}\right).\nonumber
\end{equation}
To solve $N(h)=1$ we take the parametrization:
\begin{equation}
h^I=\sqrt{\frac{2}{3}}x^I,\quad
h^M=\sqrt{\frac{2}{3}}b^M,\quad
h^4=\frac{1-b^T\bar{x}b}{\sqrt{3}||x||^2}.\nonumber
\end{equation}
where $||x||^2=\eta_{IJ}x^I x^J$ and $b^T\bar{x}b=b^M x^I \gamma_{IMN} b^N$.
\subsubsection{$SU(2)\times U(1)$ gauging}
This is the smallest member of the Magical Jordan family that admits $SU(2)\sim SO(3)$ gauging. Here we will gauge a $SU(2)\times U(1)$ subgroup of the isometry group $SL(3,\mathbb{C})$. The vector fields $A_\mu^1, A_\mu^2, A_\mu^3$ will be used to gauge $SU(2)$ and the vector $A_\mu^0$ will be the $U(1)$ gauge field. The vector fields $A_\mu^M$ are charged under $SU(2)\times U(1)$ and must be dualized to tensor fields. The vector field $A_\mu^4$ is a spectator vector field. The dualization of the vector fields to tensor fields introduces the scalar potential \cite{Gunaydin:2000ph}
\begin{equation}
 P^{(T)}=\frac{1}{8}b^M \bar{x}_{MP} \Omega^{PR} \bar{x}_{RS} \Omega^{ST}  \bar{x}_{TN} b^N\label{magicso3pt}
\end{equation}
where $\bar{x}_{MN}=\gamma_{IMN}x^I$ and the symplectic invariant matrix is
\begin{equation}
\Omega_{PR}=\left(\begin{array}{rrrr}
                    0&1&0&0\\-1&0&0&0\\
                    0&0&0&1\\0&0&-1&0\end{array}\right).\nonumber
\end{equation}
The gauge fields $A_\mu^1, A_\mu^2, A_\mu^3$ can be used to simultaneously gauge $SU(2)_R$ and this gauging leads to the potential term
\begin{equation}
 P^{(R)}=6 ||x||^2.
\end{equation}
The total potential $P^{(5)}= P^{(T)}+P^{(R)}$ does not admit any ground states. However $P^{(T)}$ itself does admit a Minkowski ground state at $b^M=0$. Using the $SU(2)$ symmetry, one can take rotate the fields such that $x^2=x^3=0$ at the critical point. With this choice the eigenvalues of the Hessian of the potential are found to be\footnote{One can arrive at the same result by doing a non-compact $SO(2,1)\times U(1)$ (without $R$-symmetry) gauging.}
\begin{equation}
 \{0,0,0,0,2 (x^0-x^1)^3,2 (x^0-x^1)^3,2 (x^0+x^1)^3,2 (x^0+x^1)^3\}\nonumber
\end{equation}
and it is easy to see that the ground state is a minimum in the region \hbox{$||x||^2>0,$} $x^0>0$ only.\\[6pt]

In the above model, one can simultaneously gauge $U(1)_R$ instead of the full $SU(2)_R$ by taking a linear combination of $A_\mu^0$ and $A_\mu^4$ as $U(1)_R$ gauge field. $P^{(T)}$ is still given by (\ref{magicso3pt}), but $P^{(R)}$ now is
\begin{equation}
P^{(R)}=-2V_0\left\{V_0 ||x||^2 + \sqrt{2} V_4 \left(2 x^0 \frac{1-b^T\bar{x}b}{||x||^2} - b^T b\right)\right\}\label{magicCprso2}
\end{equation}
where $b^T b=b^M b^N \delta_{MN}$. There are three ways of making the $b^M$-derivatives of the total potential $P^{(5)}=P^{(T)}+\lambda P^{(R)}$ vanish.\\
\underline{Case 1: $b^M=0$}\\
The $x^I$-derivatives of the potential become
\begin{equation}
 \begin{array}{rcl}
  P^{(5)}_{,0}&=&-2\lambda V_0 \left(2 V_0 x^0 -\frac{2\sqrt{2} V_4 [2 (x^0)^2 - ||x||^2]}{||x||^4}\right),\\
P^{(5)}_{,a}&=&4\lambda V_0 x^a \left(V_0-\frac{2\sqrt{2}V_4 x^0}{||x||^4}\right),\qquad a=1,2,3.
 \end{array}
\end{equation}
These vanish if one sets $x^a=0$ and $\sqrt{2} V_4 = V_0 (x^0)^3$. The value of the potential at this critical point is
 \begin{equation}P^{(5)}\arrowvert_{\varphi_C}=-6 \lambda (V_0)^2 (x^0)^2.\end{equation}
which is negative definite, hence the critical point is an Anti-deSitter ground state. The Hessian of the potential evaluated at the critical point is given by
\begin{equation}\begin{array}{rcl}
 \partial\partial P^{(5)}\arrowvert_{\varphi_C}&=\text{diag}&\left(-12\lambda (V_0)^2, -4\lambda (V_0)^2, -4\lambda (V_0)^2, -4\lambda (V_0)^2, \right. \\
&&2 (1-2\lambda (V_0)^2) (x^0)^3, 2 (1-2\lambda (V_0)^2) (x^0)^3,\\
&&\left.  2 (1-2\lambda (V_0)^2) (x^0)^3, 2 (1-2\lambda (V_0)^2) (x^0)^3\right).
\end{array}\end{equation}
Depending on the choice of $V_0$ and $V_4$ this can be either an Anti-deSitter maximum or saddle point.  

Another way of making these derivatives vanish is to set $V_0=0$. This case was already covered before. It lead to a Minkowski minimum with broken supersymmetry (unless also $V_4=0$).\\
\underline{Case 2: $b^5=b^6\text{ and } b^7=b^8$}\\
Using the $SU(2)$ invariance, the scalar fields can be rotated such that $x^2=x^3=0$. The first derivatives of the potential vanish with
\begin{equation}
 \begin{array}{rcl}
  2\sqrt{2} \lambda V_0 V_4 &=& (x^0 - x^1)^2 (x^0 + x^1),\\
 2\lambda (V_0)^2 &=&\displaystyle  \frac{x^0-3 x^1}{x^0+x^1},\\[7pt]
x^1 &=&-[(b^6)^2+(b^8)^2](x^0-x^1)^2.
 \end{array}
\end{equation}
Given a set of $V_I$'s, the values for $x^0$, $x^1$ and $(b^6)^2+(b^8)^2$ are uniquely determined by these equations. The value of the potential at the critical point is
\begin{equation}P^{(5)}\arrowvert_{\varphi_C}=-3 (x^0-x^1)^2\end{equation}
and this correspond to an Anti-deSitter vacuum. Considering the fact that we already used two thirds of the gauge freedom by choosing $x^2=x^3=0$, we conclude that this actually is a three-parameter family of ground states. Using Mathematica, we found that this is a maximum of the total potential.\\
\underline{Case 2: $b^5=-b^6\text{ and } b^7=-b^8$}\\
This is very similar to the last case. We again use the $SU(2)$ invariance to set $x^2=x^3=0$. The first derivatives of the potential vanish with
\begin{equation}
 \begin{array}{rcl}
  2\sqrt{2} \lambda V_0 V_4 &=& (x^0 - x^1) (x^0 + x^1)^2,\\
2\lambda (V_0)^2 &=&\displaystyle   \frac{x^0+3 x^1}{x^0-x^1},\\[7pt]
x^1 &=&[(b^6)^2+(b^8)^2](x^0+x^1)^2.
 \end{array}
\end{equation}
 and again, this is a three-parameter family of Anti-deSitter ground states, with the value of the potential at the critical points is
\begin{equation}P^{(5)}\arrowvert_{\varphi_C}=-3 (x^0+x^1)^2.\end{equation}
\subparagraph{With hypermultiplets:}
One can add a universal hypermultiplet to the theory and gauge simultaneously the subgroup $SU(2)\times U(1)$ of the hyperscalar isometry group $SU(2,1)$ together with the $SU(2)_R$. The total potential $P^{(5)}=P^{(T)}+P^{(R)}+P^{(H)}$ is then given by
\begin{equation}
 \begin{array}{rcl}
 P^{(T)}&=&\frac{1}{8}b^M \bar{x}_{MP} \Omega^{PR} \bar{x}_{RS} \Omega^{ST}  \bar{x}_{TN} b^N,\\
P^{(R)}&=&-4 C^{IJK} \vec{P}_I \vec{P}_J h_K,\\
P^{(H)}&=&2 \mathcal{N}_{iA} \mathcal{N}^{iA}
\end{array}
\end{equation}
with
\begin{equation}
 \mathcal{N}^{iA}=\frac{\sqrt{6}}{4}h^I K_I^X f_X^{iA}\nonumber
\end{equation}
where we defined (\ref{Tvex})
\begin{equation}
 K_0^X = T_8^X,\qquad  K_1^X = T_1^X,\qquad  K_2^X = T_2^X,\qquad  K_3^X = T_3^X.
\end{equation}
This theory does not admit any ground states. One can gauge an additional $SO(1,1)_H$ symmetry. This type of gauging admits stable and unstable deSitter vacua. But this type of calculation has been done various times in the last section and therefore we skip it here. 

\subsubsection{$SO(2,1)\times U(1)$ gauging}
We will gauge a $SO(2,1)\times U(1)$ subgroup of the isometry group $SL(3,\mathbb{C})$. The vector fields $A_\mu^0, A_\mu^1, A_\mu^2$ will be used to gauge $SO(2,1)$ and the vector $A_\mu^3$ will be the $U(1)$ gauge field. The vector fields $A_\mu^M$ are charged under $SO(2,1)\times U(1)$ and must be dualized to tensor fields. The vector field $A_\mu^4$ is a spectator vector field. The dualization of the vector fields to tensor fields introduces the scalar potential \cite{Gunaydin:2000ph}
\begin{equation}
 P^{(T)}=\frac{1}{8}b^M \bar{x}_{MP} \Omega^{PR} \bar{x}_{RS} \Omega^{ST}  \bar{x}_{TN} b^N\label{magicso21pt}
\end{equation}
where $\bar{x}_{MN}=\gamma_{IMN}x^I$ and the symplectic invariant matrix is
\begin{equation}
\Omega_{PR}=\left(\begin{array}{rrrr}
                    0&1&0&0\\-1&0&0&0\\
                    0&0&0&1\\0&0&-1&0\end{array}\right).\nonumber
\end{equation}
$P^{(T)}$ itself admits a Minkowski ground state as in the last case. In this model, one can gauge the $U(1)_R\subset SU(2)_R$ symmetry by taking a linear combination of $A_\mu^0$ and $A_\mu^4$ as $U(1)_R$ gauge field. A scalar potential 
\begin{equation}
P^{(R)}=2V_3\left\{V_3 ||x||^2 + \sqrt{2} V_4 \left(-2 x^3 \frac{1-b^T\bar{x}b}{||x||^2} + b^T \gamma_3 b\right)\right\}\label{magicCprso21}
\end{equation}
is introduced. The only critical points of the total potential $P^{(5)}=P^{(T)}+\lambda P^{(R)}$ are found at $b^M=0$ by setting $V_3=0$,i.e. by turning off the $U(1)_R$ potential. These are four-parameter family of Minkowski ground states and they are minima only in the region $||x||^2>0,\quad x^0>0$. Supersymmetry is broken unless also $V_4=0$.\\
\subsection{Summary}
The gaugings of certain theories in the generic Jordan family can be reproduced in the Magical Jordan family, provided there are enough vector fields to do the respective gaugings. The stability of the ground states of these theories still needs to be checked and in some cases the stability puts constraints on the gauge parameters. In this section we reproduced the Minkowski and Anti-Desitter ground states for $SO(2)\times U(1)_R$ gauging and the deSitter ground states for $SO(1,1)\times U(1)_R$ gauging that were already found in the generic Jordan family case. In addition to the existing ground states, we encountered other ground states that are special to the Magical Jordan family case, such as deSitter and Anti-deSitter saddle points and curves. Although we did not do a complete analysis, we can conclude that the Magical Jordan family theories are richer than the generic Jordan family theories in the numbers and properties of ground states. 

The compact non-Abelian $SU(2)\times U(1)$ gauging leads to Minkowski vacua; and also Anti-deSitter vacua when accompanied by a simultaneous $U(1)_R$ symmetry gauging. However the simultaneous $SU(2)_R$ gauging does not admit any critical points, even after including hypers in the model. The model with non-compact non-Abelian $SO(2,1)\times U(1)$ gauging has Minkowski vacuum, but doing a simultaneous $U(1)_R$ gauging results in a theory with no ground states.

The other members of the Magical Jordan family ($\mathcal{M}=SU*(6)/USp(6)$ and $\mathcal{M}=E_{6(-26)}/F_4$) have a very similar structure to the above theories and contain them as subsectors. Although they admit gaugings of bigger subgroups, such as $SO(m+1)$ or $SO(m,1)$ with $m\geq 3$, the form of the scalar potentials corresponding to the $SO(3)$ and $SO(2,1)$ gaugings of the above model suggests that it is not likely for the bigger members of the Magical Jordan family, subject to $SO(m+1)$ or $SO(m,1)$ gaugings, to have ground states of different nature than the ones found in this section.

\section{Generic non-Jordan Family}
 The scalar manifold $\mathcal{M}=SO(1,\tilde{n})/SO(\tilde{n})$ can be described by the hyper surface $N(h)=C_{\tilde{I}\tilde{J}\tilde{K}}h^{\tilde{I}}h^{\tilde{J}}h^{\tilde{K}}=1$ of the cubic polynomial
\begin{equation}
 N(h)=\frac{3\sqrt{3}}{2\sqrt{2}}\left(\sqrt{2} h^0 (h^1)^2 - h^1 \left[(h^2)^2 +...+(h^{\tilde{n}})^2\right]\right).
\end{equation}
The non-vanishing $C_{IJK}$'s are
\begin{equation}
 C_{011}=\frac{\sqrt{3}}{2},\qquad C_{022}=C_{033}=...=C_{0\tilde{n}\tilde{n}}=-\frac{\sqrt{3}}{2\sqrt{2}}\nonumber
\end{equation}
and their permutations. To solve $N(h)=1$ we take the parametrization
\begin{equation}\begin{array}{rcl}
 h^0&=&\displaystyle\sqrt{\frac{2}{3}}\left(\frac{1}{\sqrt{2}(\varphi^1)^2}+\frac{1}{\sqrt{2}}\varphi^1 \left[(\varphi^2)^2+...+(\varphi^{\tilde{n}})^2\right]\right),\\[8pt]
h^1&=&\displaystyle\sqrt{\frac{2}{3}} \varphi^1,\\[8pt]
h^a&=&\displaystyle\sqrt{\frac{2}{3}} \varphi^1 \varphi^a,\qquad a=2,...,\tilde{n}.\nonumber
\end{array}\end{equation}
In contrast to the Jordan families, $C^{\tilde{I}\tilde{J}\tilde{K}}$'s are no longer constant or equal to $C_{\tilde{I}\tilde{J}\tilde{K}}$'s. The scalar field dependent $C^{\tilde{I}\tilde{J}\tilde{K}}$ are defined as
\begin{equation}
 C^{\tilde{I}\tilde{J}\tilde{K}}=\quad \stackrel{o}{a}^{\tilde{I}\tilde{I'}} \stackrel{o}{a}^{\tilde{J}\tilde{J'}} \stackrel{o}{a}^{\tilde{K}\tilde{K'}} C_{\tilde{I'}\tilde{J'}\tilde{K'}}\nonumber
\end{equation}
where the inverse of the vector field metric $\stackrel{o}{a}_{\tilde{I}\tilde{J}}$ is given by $\stackrel{o}{a}^{\tilde{I}\tilde{J}}=h^{\tilde{I}} h^{\tilde{J}} + h^{\tilde{I}}_{\tilde{x}} h^{\tilde{J}}_{\tilde{y}} g^{\tilde{x}\tilde{y}}$. For the symmetric non-Jordan family, the scalar field metric $g_{\tilde{x}\tilde{y}}$ is diagonal
\begin{equation}
 g_{\tilde{x}\tilde{y}}=\text{diag }[\frac{3}{(\varphi^1)^2},(\varphi^1)^3,...,(\varphi^1)^3]\nonumber
\end{equation}
which is positive definite for $\varphi^1>0$.
\subsection{Maxwell-Einstein supergravity}
\subsubsection{No $R$-symmetry gauging}
We add one hypermultiplet to the theory and gauge a non-compact $SO(1,1)_H$ symmetry of the hyperscalar manifold. As the $SO(1,1)$ gauge field, we take a linear combination $W_I A^I_\mu$ of all the vectors in the theory. The potential is given by
\begin{equation}
 P^{(5)}_{TOT}=P^{(H)}=2 \mathcal{N}_{iA} \mathcal{N}^{iA}
\end{equation}
where $\mathcal{N}^{iA}=\frac{\sqrt{6}}{4}(W_I h^I) T_4^X f_X^{iA}$. The only way to make the first derivatives of the potential vanish at the base point of the hyperscalar manifold without making the potential itself vanish is to set
\begin{equation}\begin{array}{rcl}
 W_1 &=&\displaystyle   \frac{W_0 \left(2+(\varphi^1)^3\left[(\varphi^2)^2+...+(\varphi^{\tilde{n}})^2\right]\right)}{\sqrt{2} (\varphi^1)^3},\\[9pt]
W_a&=&-\sqrt{2} \varphi^a W_0,\qquad a=2,...,\tilde{n}.
\end{array}\end{equation}
The coordinates of the critical point is entirely determined by $W_I$'s. The value of the potential at the critical point becomes
\begin{equation}
  P^{(5)}_{TOT}\arrowvert_{\varphi^C}=\frac{9}{4}\frac{(W_0)^2}{(\varphi^1)^4}
\end{equation}
and the Hessian of the potential at the critical point is given by
\begin{equation}
 \partial\partial P^{(5)}_{TOT}\arrowvert_{\varphi^C}=\text{diag }[\frac{9 (W_0)^2}{(\varphi^1)^6},\underbrace{\frac{3 (W_0)^2}{\varphi^1},...,\frac{3 (W_0)^2}{\varphi^1}}_{\tilde{n}-1 \text{ times}},0,\frac{9 (W_0)^2}{2(\varphi^1)^4},\frac{9 (W_0)^2}{4(\varphi^1)^4},\frac{9 (W_0)^2}{4(\varphi^1)^4}]\nonumber
\end{equation}
which is semi-positive definite in the physically relevant region $\varphi^1>0$, therefore the critical point is a stable deSitter vacuum. We already had many examples of having $SO(1,1)_H$ gauging mixed with other gaugings in the generic Jordan family section. Similar analysis for the non-Jordan family shows that it is possible to obtain deSitter ground states with other gauge groups that include $SO(1,1)_H$. Therefore we omit the results for $K\times SO(1,1)_H$ gaugings of generic non-Jordan family theories.
\subsubsection{$SU(2)_R$ symmetry gauging}
This calculation was done in \cite{Gunaydin:2000ph}. Let us briefly quote their results. The vectors $A_\mu^2, A_\mu^3, A_\mu^4$ are chosen as the $SU(2)$ gauge fields. This group rotates $h^2, h^3, h^4$ together bu the other scalars are not charged under the action of this $SU(2)$, therefore no tensor fields need to be introduced. The scalar potential (\ref{su2r}) becomes
\begin{equation}P^{(5)}_{TOT}=P^{(R)}=-\frac{1}{2}(\varphi^1)^2\left[(\varphi^2)^2+(\varphi^3)^2+(\varphi^4)^2\right]+\frac{3}{2\varphi^1}
\end{equation}
It's easy to verify that this potential does not have any critical points.
\subsubsection{$U(1)_R$-symmetry gauging}
This calculation was done in \cite{smet} for $\tilde{n}=3$. Let us trivially generalize their results to arbitrary $\tilde{n}$. A linear combination $V_I A^I_\mu$ of all the vectors in the theory is taken as the $U(1)_R$ gauge field. The scalar potential (\ref{u1r}) is given by
\begin{equation}\begin{array}{rl}
 P^{(5)}_{TOT}=P^{(R)}=\frac{1}{\varphi^1}&\{-2\sqrt{2} V_0 V_1 +2 |V|^2 \\ &- (\varphi^1)^3\left[V_0 ||\tilde{\varphi}||^2 +\sqrt{2} (V_1+V_2 \varphi^2+...+V_{\tilde{n}}\varphi^{\tilde{n}})\right]^2\}
\end{array}\end{equation}
where we defined $|V|^2=(V_2)^2+...+(V_{\tilde{n}})^2$ and $||\tilde{\varphi}||^2 = (\varphi^2)^2 + ... + (\varphi^{\tilde{n}})^2
$. The only way to make the first derivatives of the potential vanish without making the potential itself vanish is to set
\begin{equation}\begin{array}{rcl}
 V_1 &=&\displaystyle   \frac{V_0 \left(2+(\varphi^1)^3 ||\tilde{\varphi}||^2\right)}{\sqrt{2} (\varphi^1)^3},\\[8pt]
V_a&=&\displaystyle  -\sqrt{2} \varphi^a V_0,\qquad a=2,...,\tilde{n}.
\end{array}\end{equation}
The coordinates of the critical point is entirely determined by $V_I$'s. The value of the potential at the critical point becomes
\begin{equation}
  P^{(5)}_{TOT}\arrowvert_{\varphi^C}=-\frac{24 (\varphi^1)^2 (V_1)^2}{(2+(\varphi^1)^3 ||\tilde{\varphi}||^2)^2}
\end{equation}
and the Hessian of the potential at the critical point is given by
\begin{equation}
 \partial\partial P^{(5)}_{TOT}\arrowvert_{\varphi^C}=\text{diag }[-3\tilde{A},\underbrace{-\tilde{A} (\varphi^1)^5,...,-\tilde{A} (\varphi^1)^5}_{\tilde{n}-1 \text{ times}}]\nonumber
\end{equation}
where $\tilde{A}=\frac{(V_1)^2}{(2+(\varphi^1)^3 ||\tilde{\varphi}||^2)^2}>0$. Thus the critical point is an Anti-deSitter maximum.
\subsection{Yang-Mills/Einstein supergravity with tensor coupling}
The Lagrangian of the theory is not invariant under full the isometry group $SO(1,\tilde{n})$, but it is invariant under the subgroup $G=[SO(\tilde{n}-1,1)\times SO(1,1)]\ltimes T_{\tilde{n}-1}$, where $T_{\tilde{n}-1}$ is the group of translations in an $\tilde{n}-1$ dimensional Euclidean space. Having a closer look at $N$, we see that the subgroup $SO(1,1)$ cannot be gauged because all the vector fields are charged under it and there are no vector field to be used as the gauge field. Only the gauging of the subgroup $SO(2)\subset SO(\tilde{n}-1,1)$ will result in a potential term due to dualization of the vector fields to tensor fields.
\subsubsection{No $R$-symmetry gauging}
The group $SO(2)$ rotates $h^2$ and $h^3$ into each other and therefore acts nontrivially on the vector fields $A^2_\mu$ and $A^3_\mu$. These fields must be dualized to tensor fields. The field $A_\mu^1$ is chosen as the $SO(2)$ gauge field. The index $\tilde{I}$ is decomposed as \[\tilde{I}=(I,M)\] where $I,J,K=0,1,4,5,...,\tilde{n}$ and $M,N,P=2,3$. The scalar potential (\ref{pt}) is found as
\begin{equation}
  P^{(5)}_{TOT}=P^{(T)}=\frac{(\varphi^1)^5}{8}\left[(\varphi^2)^2+(\varphi^3)^2\right].
\end{equation}
This potential has an $\tilde{n}-2$ parameter family of Minkowski minima at $\varphi^2=\varphi^3=0$.
\subsubsection{$SU(2)_R$-symmetry gauging}
The vector fields $A_\mu^4, A_\mu^5, A_\mu^6$ are chosen as $SU(2)_R$ gauge fields, whereas $A_\mu^1$ will be used to gauge $SO(2)$. The vectors  $A_\mu^2, A_\mu^3$ transform nontrivially under $SO(2)$, therefore they are dualized to tensor fields. The total potential $P^{(5)}_{TOT}=P^{(T)}+\lambda P^{(R)}$ is given by
\begin{equation}
 \begin{array}{rcl}
 P^{(T)}&=&\displaystyle\frac{(\varphi^1)^5}{8}\left[(\varphi^2)^2+(\varphi^3)^2\right],\\[6pt]
 P^{(R)}&=&-\displaystyle\frac{1}{2}(\varphi^1)^2\left[(\varphi^4)^2+(\varphi^5)^2+(\varphi^6)^2\right]+\displaystyle\frac{3}{2\varphi^1}.
\end{array}
\end{equation}
It is easy to verify that the total potential does not have any critical points.
\subsubsection{$U(1)_R$-symmetry gauging}
As in the last model, $A_\mu^1$ is the $SO(2)$ gauge field and because the vectors  $A_\mu^2, A_\mu^3$ transform nontrivially under $SO(2)$, they are dualized to tensor fields. A linear combination $V_I A^I_\mu, I=0,1,4,5,...,\tilde{n}$ of vector fields is used as the $U(1)_R$ gauge field. The total potential $P^{(5)}_{TOT}=P^{(T)}+\lambda P^{(R)}$ is given by
 \begin{equation}
\begin{array}{rcl}
 P^{(T)}&=&\displaystyle  \frac{(\varphi^1)^5}{8}\left[(\varphi^2)^2+(\varphi^3)^2\right],\\[6pt]
 P^{(R)}&=&\displaystyle  \frac{1}{\varphi^1}\{-2\sqrt{2} V_0 V_1 +2 |V|^2 \\ &&\text{ }\quad- (\varphi^1)^3\left[V_0 ||\tilde{\varphi}||^2 +\sqrt{2} (V_1+V_4 \varphi^4+...+V_{\tilde{n}}\varphi^{\tilde{n}})\right]^2\},
\end{array}
\end{equation}
where $|V|^2=(V_4)^2+...+(V_{\tilde{n}})^2$ and $||\tilde{\varphi}||^2 = (\varphi^2)^2 + ... + (\varphi^{\tilde{n}})^2
$. The first derivatives of the potential are given by
 \begin{equation}
\begin{array}{rcl}
 \partial_{\varphi^1} P^{(5)}_{TOT} &=& \frac{5}{8} (\varphi^1)^4 [(\varphi^2)^2+(\varphi^3)^2]+\frac{4\sqrt{2}\lambda V_0 V_1 - 2 |V|^2}{(\varphi^1)^2}-2\lambda \varphi^1 A^2,\\
\partial_{\varphi^a} P^{(5)}_{TOT} &=& \frac{1}{4} (\varphi^1)^2 \varphi^a \left((\varphi^1)^3 - 16\lambda V_0 A\right),\qquad a=2,3\\
\partial_{\varphi^b} P^{(5)}_{TOT} &=& -2\sqrt{2} \lambda (\varphi^1)^2 (\sqrt{2} \varphi^b V_0 + V_b) A,\qquad b=4,...,\tilde{n},\label{derivnonjord}
\end{array}
\end{equation}
where
\begin{equation}
 A=||\tilde{\varphi}||^2 V_0 +\sqrt{2} (V_1 + \varphi^4 V_4 +...+ \varphi^{\tilde{n}} V_{\tilde{n}}).
\end{equation}
There are three ways of making these expressions vanish. 
\subparagraph{\underline{Case 1: $\varphi^a = A = 0$}}
In this case we have
\begin{equation}
 P^{(5)}_{TOT}\arrowvert_{\varphi^C} = -\varphi^1 \partial_{\varphi^1} P^{(5)}_{TOT}\arrowvert_{\varphi^C}
\end{equation}
which means the potential vanishes at the critical point.
\subparagraph{\underline{Case 2: $\varphi^a = 0, A \neq 0$}}
In this case one must have $V_b = -\sqrt{2} \varphi^b V_0$ to make the third expression in (\ref{derivnonjord}) vanish. Plugging this into the first expression and setting it equal to zero, one finds the conditions
\begin{equation}
 V_1=\frac{||\tilde{\varphi}||^2 V_0}{\sqrt{2}},\quad\text{or}\quad V_1=\frac{2+(\varphi^1)^3 ||\tilde{\varphi}||^2 V_0}{\sqrt{2} (\varphi^1)^3}.
\end{equation}
The first of these leads to a Minkowski minimum. The second choice gives the value of the potential at the critical point as
\begin{equation}
P^{(5)}_{TOT}\arrowvert_{\varphi^C} =-\frac{12\lambda (V_0)^2}{(\varphi^1)^4}
\end{equation}
and the Hessian of the potential at the critical point is
\begin{equation}\begin{array}{rl}
 \partial\partial P^{(5)}_{TOT}\arrowvert_{\varphi^C}=\text{diag}&\displaystyle[-\frac{24\lambda (V_0)^2}{(\varphi^1)^6},\frac{(\varphi^1)^5}{4}-\frac{8\lambda (V_0)^2}{\varphi^1},\frac{(\varphi^1)^5}{4}-\frac{8\lambda (V_0)^2}{\varphi^1},\\[12pt] & \underbrace{-\frac{8\lambda (V_0)^2}{\varphi^1},...,-\frac{8\lambda (V_0)^2}{\varphi^1}}_{(\tilde{n}-3)\text{ times}}]
\end{array}\end{equation}
which means the critical point can be a maximum or a saddle point depending on the choice of $V_I$'s.
\subparagraph{\underline{Case 3: $\varphi^a \neq 0$}}
In this case one must have $V_b = -\sqrt{2} \varphi^b V_0$ together with $(\varphi^1)^3 = 16\lambda V_0 A$. Plugging these in the first equation in (\ref{derivnonjord}), solving this for $V_1$ one finds the value of the potential at the critical point as
\begin{equation}
P^{(5)}_{TOT}\arrowvert_{\varphi^C} =-\frac{(\varphi^1)^5\left(-3 (\varphi^1)^3 + 32 \lambda (5+\lambda) [(\varphi^2)^2+(\varphi^3)^2] (V_0)^2\right)}{256 \lambda^2 (V_0)^2}.
\end{equation}
which might correspond to deSitter or Anti-deSitter, depending on the choice of $V_I$'s. It was shown in \cite{smet} that the deSitter solution is a saddle point when $\tilde{n}=3$. The calculation for the stability of the solutions are tedious but using Mathematica, we confirmed that the deSitter solutions are saddle points for any $\tilde{n}$ and we showed that the Anti-deSitter solutions are either maxima or saddle points, again depending on the choice of $V_I$'s. 
\subsection{Summary}
$SU(2)_R$ gauging does not lead to any critical points, even with the addition of tensors; whereas the model with pure $U(1)_R$ gauging has Minkowski and AdS critical points. The only way of adding tensors to the theory is done by gauging the $SO(2)$ subgroup of the isometry group. Pure $SO(2)$ gauging leads to Minkowski minima. $U(1)_R\times SO(2)$ gauging has Minkowski, dS and AdS critical points. The dS solution is always unstable but the AdS solution can be made stable by properly choosing $V_I$'s (c.f. \cite{Townsend:1984iu}). Coupling hypers to the theory and gauging $SO(1,1)_H$ leads to stable deSitter vacua as in the generic Jordan case.

\section{Conclusions}
In this paper, after reviewing the ground state solutions of the $5D, \mathcal{N}=2$ supergravity theories with symmetric scalar manifolds that had been discovered earlier, we studied the vacua of the gauged $5D, \mathcal{N}=2$ supergravity theories that had not been discussed in the literature. Consistent with earlier results, in the absence of hypers, we showed that all the generic Jordan family, the Magical Jordan family and the generic non-Jordan family theories admit stable Anti-deSitter vacua, whereas only the theories of the first two families admit stable deSitter vacua and all the above families have unstable deSitter and Anti-deSitter ground states.  

For the generic Jordan family, the only gauge groups $K$ that lead to the introduction  of tensor fields are the Abelian groups $SO(2)$ and $SO(1,1)$. The former leads to supersymmetric Minkowski ground states only, unless accompanied by a simultaneous $U(1)_R$ gauging. With $U(1)_R$ gauging one can obtain nonsupersymmetric Minkowski ground states, and moreover, there are supersymmetric and nonsupersymmetric Anti-deSitter critical points resulting from the combined scalar potential of the $SO(2)$ and $U(1)_R$ gaugings. The $SO(1,1)$ gauging, on the other hand, breaks the supersymmetry and leads to stable deSitter vacua by a simultaneous $R$-symmetry ($SU(2)_R$ or $U(1)_R$) gauging. Pure $SO(1,1)$ gauging does not lead to any critical points. It is interesting to observe that whereas the $SO(1,1)\times SU(2)_R$ gauging has stable deSitter vacua, its compact counterpart, namely $SO(2)\times SU(2)_R$ gauging admits Minkowski vacua only. We also note that some of the stable deSitter models we studied; such as the one with $SO(1,1)\times SU(2)_R$ gauging and no hypermultiplets, or the one with $SO(2)\times SO(1,1)_H$ gauging, which has one hypermultiplet; have critical points, for which the Hessians of the potential evaluated at these points have zero eigenvalues. This is related to the fact that the potential has a family of critical points rather than a single point and therefore it has flat direction(s) at these points.

We showed that it is possible to embed certain generic Jordan family models into the Magical Jordan family theories, provided that there are sufficient number of vector fields in the Magical theory to do the respective gauging. However, in some cases the stability puts additional constraints on the gauge parameters. In these models, we encountered other critical points than the ones obtained in the generic case, such as deSitter and Anti-deSitter saddle points and curves. These are special to the Magical Jordan family. Although we found numerous critical points of these models, we could not do a complete analysis due to the complexity of the Magical Jordan family theories. In addition to this, coupling a hypermultiplet and gauging a subgroup of its scalar manifold might lead to nontrivial critical points that are beyond those found in this paper. These are left as open questions for future investigation. 

Other than the embeddings of the generic Jordan family cases, one can gauge non-Abelian subgroups of the isometry groups of the Magical theories and dualize non-trivially charged vector fields to tensor fields which yields additional contributions to the scalar potential. The compact $SO(3)\times U(1)$ gauging leads to a Minkowski vacuum. A simultaneous $SU(2)_R$ gauging leads to a theory with no critical points whereas a simultaneous $U(1)_R$ gauging has an Anti-deSitter solution. On the other hand, the non-compact non-Abelian $SO(2,1)\times U(1)$ gauging only leads to a Minkowski ground state and adding a simultaneous $R$-symmetry gauging results in a theory with no ground states.

For the generic non-Jordan family, the model with the full $R$-symmetry gauged does not have any critical points, even after adding tensor coupling. The pure $U(1)_R$ gauging leads to Minkowski and Anti-deSitter ground states. Tensor coupling to these models can only be achieved by doing a compact $SO(2)$ gauging. A simultaneous $SO(2)\times U(1)_R$ gauging results in Minkowski, Anti-deSitter and deSitter ground states. The deSitter solutions are found to be unstable whereas the Anti-deSitter solutions can be made stable by proper choices of the parameters $V_I$ that define the linear combination of the vector fields that is used as the $U(1)_R$ gauge field.

We also added a universal hypermultiplet to the models we considered and investigated the potentials coming from the gauging of the hyper isometries. For the generic Jordan family, we showed that a simultaneous compact $U(1)_H$ gauging does not change the sign of the potential at the existing critical points of the models that the hypermultiplet is added to, but a non-compact $SO(1,1)_H$ gauging generally leads to deSitter vacua. It is interesting to see that the $SO(1,1)$ gaugings of both real and hyperscalar isometries help finding deSitter ground states. This result is not limited to the generic Jordan family and applies to the other families.

{\bf Acknowledgement:} The author thanks Murat G\"unaydin for suggesting the investigation that is the subject of this paper and for his guidance. This work was supported in part by the National Science Foundation under grant
number  PHY-0555605. Any opinions, findings and conclusions or
recommendations expressed in this material are those of the author
and do not necessarily reflect the views of the National Science
Foundation.

\appendix
\section{The ``Very Special Geometry''}

\setcounter{equation}{0}
The bosonic sector of the $5D, \mathcal{N}=2$ gauged Yang-Mills-Einstein supergravity\footnote{For the full Lagrangean, see \cite{Gunaydin:Vacua},\cite{Ceresole:2000jd}} coupled to tensor- and hypermultiplets is described by the Lagrangean (with metric signature $(-++++)$) \cite{Gunaydin:1999zx},\cite{Ellis:2001xd},\cite{Ceresole:2000jd}
\begin{equation}\begin{array}{rcl}
\hat{e}^{-1}\mathcal{L}_{bosonic}^{\mathcal{N}=2}&=&\displaystyle-\frac{1}{2}R-\frac{1}{4}\stackrel{o}{a}_{\tilde{I}\tilde{J}}\mathcal{H}^{\tilde{I}}_{\mu\nu}\mathcal{H}^{\tilde{J}\mu\nu}-\frac{1}{2}g_{XY}\mathcal{D}_\mu q^X \mathcal{D}^\mu q^Y\\[7pt]
&&-\displaystyle\frac{1}{2}g_{\tilde{x}\tilde{y}} \mathcal{D}_\mu \varphi^{\tilde{x}} \mathcal{D}^\mu \varphi^{\tilde{y}} +\frac{\hat{e}^{-1}}{6\sqrt{6}} C_{IJK} \epsilon^{\mu\nu\rho\sigma\tau} F^I_{\mu\nu} F^J_{\rho\sigma} A^K_\tau\\[7pt]
&&\displaystyle+\frac{\hat{e}^{-1}}{4 g} \epsilon^{\mu\nu\rho\sigma\tau} \Omega_{MN} B^M_{\mu\nu} \mathcal{D}_\rho B^N_{\sigma\tau} -\mathcal{V}(\varphi, q).
\end{array}\end{equation}
Here, non-Abelian field strengths $\mathcal{F^I}_{\mu\nu}\equiv  F^I_{\mu\nu} + g f^I_{JK} A^J_\mu A^K_\nu\quad (I=0,1,...,n)$ of the gauge group $K$ and the self-dual tensor fields $B^M_{\mu\nu}\quad (M=1,2,...,2m)$ are grouped together to define the tensorial quantity $\mathcal{H}^{\tilde{I}}_{\mu\nu} \equiv (\mathcal{F^I}_{\mu\nu},B^M_{\mu\nu})$ with $\tilde{I}=0,1,...,n+2m$. The potential term $\mathcal{V}(\varphi, q)$ is given by 
\begin{equation}
 \mathcal{V}(\varphi, q) = g^2 (P^{(T)}(\varphi) + \lambda P^{(R)}(\varphi, q)+\kappa P^{(H)}(q))
\end{equation}
where
\begin{equation}
 \begin{array}{rcl}
  P^{(T)} &=& 2 W_{\tilde{x}} W^{\tilde{x}}\\[4pt]
P^{(R)} &=& -4 \vec{P}\cdot \vec{P} +2 \vec{P}^{\tilde{x}}\cdot \vec{P}_{\tilde{x}}\\[4pt]
P^{(H)} &=& 2 \mathcal{N}_{X} \mathcal{N}^{X}
 \end{array}
\end{equation}
and $\lambda=g_R^2/g^2$ , $\kappa=g_H^2/g^2$. The quantities given in the above expression are defined as
\begin{equation}
 \begin{array}{rcl}
 W_{\tilde{x}}&\equiv&\displaystyle-\frac{\sqrt{6}}{8} \Omega^{MN} h_{M\tilde{x}} h_N = \frac{\sqrt{6}}{4} h^I K_I^{\tilde{x}},\\[7pt]
\vec{P}&\equiv&h^I \vec{P}_I,\\[7pt]
\vec{P}_{\tilde{x}}&\equiv&h^I_{\tilde{x}} \vec{P}_I,\\[7pt]
\mathcal{N}^{X}&\equiv&\displaystyle\frac{\sqrt{6}}{4}h^I K_I^{X},\label{wapadef}
\end{array}
\end{equation}
where $K_I^{\tilde{x}}$ and $K_I^{X}$ are Killing vectors acting on the scalar and the hyperscalar parts of the total scalar manifold $\mathcal{M}_{scalar} = \mathcal{M}_{VS} \otimes \mathcal{M}_{Q}$; $\vec{P}_I$ are the Killing prepotentials which will be defined below; $\Omega^{MN}$ is the inverse of $\Omega_{MN}$, which is the  constant invariant anti-symmetric tensor of the gauge group $K$; and $h^I$ and $h^I_{\tilde{x}}$ are elements of the very special manifold $\mathcal{M}_{VS}$ described by the hypersurface 
\begin{equation}
 N(h)=C_{\tilde{I}\tilde{J}\tilde{K}} h^{\tilde{I}} h^{\tilde{J}} h^{\tilde{K}} = 1,\qquad \tilde{I},\tilde{J},\tilde{K} = 0,...,\tilde{n}
\end{equation}
of the $\tilde{n}+1$ dimensional space $M=\{h^{\tilde{I}} \in \mathbb{R}^{\tilde{n}+1}|N(h)=C_{\tilde{I}\tilde{J}\tilde{K}} h^{\tilde{I}} h^{\tilde{J}} h^{\tilde{K}}>0\}$ with metric 
\begin{equation}
 a_{IJ} = -\frac{1}{3}\partial_I \partial_J \text{ln} N(h).
\end{equation}
The terms $P^{(T)}$ and $P^{(H)}$ are semi-positive definite in the physically relevant region, whereas $P^{(R)}$ can have both signs. $\mathcal{M}_{VS}$ is determined completely by the totally symmetric tensor $C_{\tilde{I}\tilde{J}\tilde{K}}$. The scalar field metric on this hypersurface is the induced metric from the embedding space, which is given by
\begin{equation}
 g_{\tilde{x}\tilde{y}}=\frac{3}{2} a_{\tilde{I}\tilde{J}} h^{\tilde{I}}_{,\tilde{x}} h^{\tilde{J}}_{,\tilde{y}} \arrowvert_{N=1}=-3C_{\tilde{I}\tilde{J}\tilde{K}} h^{\tilde{I}} h^{\tilde{J}}_{,\tilde{x}} h^{\tilde{K}}_{,\tilde{y}}\arrowvert_{N=1}\label{appgxy}
\end{equation}
where $,\tilde{x}$ denotes a derivative with respect to $\varphi^{\tilde{x}}$. The definitions
\begin{equation}\begin{array}{rcl}
\stackrel{o}{a}_{\tilde{I}\tilde{J}} &\equiv& a_{\tilde{I}\tilde{J}} \arrowvert_{N=1}=-2C_{\tilde{I}\tilde{J}\tilde{K}} h^{\tilde{K}}+ 3 h_{\tilde{I}} h_{\tilde{J}},\\
h_{\tilde{I}}&\equiv& C_{\tilde{I}\tilde{J}\tilde{K}} h^{\tilde{J}} h^{\tilde{K}} = \stackrel{o}{a}_{\tilde{I}\tilde{J}} h^{\tilde{J}},\\
h^{\tilde{I}}_{\tilde{x}}&\equiv&-\sqrt{\frac{3}{2}} h^{\tilde{I}}_{,\tilde{x}},\\
h_{\tilde{I}\tilde{x}}&\equiv&\stackrel{o}{a}_{\tilde{I}\tilde{J}} h^{\tilde{J}}_{\tilde{x}}=\sqrt{\frac{3}{2}}h_{\tilde{I},\tilde{x}}
\end{array}\label{appdefs}\end{equation}
help us write the algebraic constraints of the very special geometry
\begin{equation}
 \begin{array}{rcl}
h^{\tilde{I}} h_{\tilde{I}}&=&1,\\
h^{\tilde{I}}_{\tilde{x}} h_{\tilde{I}}&=&h_{\tilde{I}\tilde{x}} h^{\tilde{I}}=0,\\
h^{\tilde{I}}_{\tilde{x}} h^{\tilde{J}}_{\tilde{y}} \stackrel{o}{a}_{\tilde{I}\tilde{J}} &=& g_{\tilde{x}\tilde{y}}.
\end{array}\label{appalco}
\end{equation}
There are also differential constraints to be satisfied:
\begin{equation}
 \begin{array}{rcl}
h_{\tilde{I} \tilde{x};\tilde{y}}&=&\sqrt{\frac{2}{3}}\left(g_{\tilde{x}\tilde{y}} h_{\tilde{I}} +T_{\tilde{x}\tilde{y}\tilde{z}} h^{\tilde{z}}_{\tilde{I}}\right),\\
h^{\tilde{I}}_{ \tilde{x};\tilde{y}}&=&-\sqrt{\frac{2}{3}}\left(g_{\tilde{x}\tilde{y}}h^{\tilde{I}}+T_{\tilde{x}\tilde{y}\tilde{z}} h^{\tilde{I}\tilde{z}}\right).
\end{array}
\end{equation}
where $';'$ is the covariant derivative using the Christoffel connection calculated from the metric $g_{\tilde{x}\tilde{y}}$ and
\begin{equation}
 T_{\tilde{x}\tilde{y}\tilde{z}}\equiv C_{\tilde{I}\tilde{J}\tilde{K}} h^{\tilde{I}}_{ \tilde{x}} h^{\tilde{J}}_{ \tilde{y}} h^{\tilde{K}}_{ \tilde{z}}.
\end{equation}
Using (\ref{appgxy}),(\ref{appdefs}) and (\ref{appalco}) one can derive
\begin{eqnarray}
 \stackrel{o}{a}_{\tilde{I}\tilde{J}}&=&h_{\tilde{I}} h_{\tilde{J}} + h_{\tilde{I}}^{ \tilde{x}} h_{\tilde{J} \tilde{x}}\\
h_{\tilde{I}}^{ \tilde{x}} h_{\tilde{J} \tilde{x}}&=& -2 C_{\tilde{I}\tilde{J}\tilde{K}} h^{\tilde{K}} + 2 h_{\tilde{I}} h_{\tilde{J}}.
\end{eqnarray}
The indices $\tilde{I},\tilde{J},\tilde{K}$ are raised and lowered by  $\stackrel{o}{a}_{\tilde{I}\tilde{J}}$ and its inverse $\stackrel{o}{a}^{\tilde{I}\tilde{J}}$. $P^{(T)}$ can now be written in a more compact form
\begin{equation}
 \begin{array}{rcl}
  P^{(T)}&=& \frac{3}{8} \Omega^{MN} \Omega^{PR} C_{MRI} h_N h_P h^I\\
&=&\frac{3\sqrt{6}}{16} \Lambda_I^{MN} h_M h_N h^I.
 \end{array}
\end{equation}
with $\Lambda^M_{IN}$ being the transformation matrices of the tensor fields under the gauge group $K$
\begin{equation}
 \Lambda_I^{MN} = \Lambda^M_{IP} \Omega^{PN}=\frac{2}{\sqrt{6}} \Omega^{MR} C_{IRP} \Omega^{PN}.
\end{equation}
 Gauging the $R$-symmetry introduces the potential term $P^{(R)}=-4 \vec{P}\cdot \vec{P} +2 \vec{P}^{\tilde{x}}\cdot \vec{P}_{\tilde{x}}$, where $\vec{P}=h^I \vec{P}_I$ and $\vec{P}_{\tilde{x}}=h^I_{\tilde{x}}\vec{P}_I$ are vectors that transform under the $R$-symmetry group that is being gauged. For the $SU(2)_R$ gauging one can take
\begin{equation}
 \vec{P}_I = \vec{e}_I\nonumber
\end{equation}
where $\vec{e}_I$ satisfy $\vec{e}_A\times \vec{e}_B=d_{AB}^{\phantom{AB}{C}} \vec{e}_C$ and $\vec{e}_A\cdot  \vec{e}_B=\delta_{AB}$ when $A,B,C$ are the $SU(2)_R$ adjoint indices ($d_{AB}^{\phantom{AB}{C}}$ are the $SU(2)$ structure constants); and $\vec{e}_I=0$ otherwise. With this convention and the use of (\ref{appdefs}) and (\ref{appalco}) the potential term simplifies to
\begin{equation}
P^{(R)}=-4 C^{AB\tilde{K}}\delta_{AB} h_{\tilde{K}}.
\end{equation}
If the $U(1)_R$ subgroup of $SU(2)_R$ is being gauged one can take
\begin{equation}
 \vec{P}_I = V_I \vec{e},\nonumber
\end{equation}
where $\vec{e}$ is an arbitrary vector in the $SU(2)$ space and $V_I$ are some constants that define the linear combination of the vector fields $A^I_\mu$ that is used as the $U(1)_R$ gauge field
\begin{equation}
A_\mu [U(1)_R]=V_I A^I_\mu.\nonumber
\end{equation}
The potential term then can be written as
\begin{equation}
P^{(R)}=-4C^{IJ\tilde{K}}V_I V_J h_{\tilde{K}}.
\end{equation}

If tensors are coupled to the theory the $V_I$ have to be constrained by
\begin{equation}
V_I f^I_{JK}=0\nonumber
\end{equation}
with $f^I_{JK}$ being the structure constants of $K$. When the target manifold $\mathcal{M}_{VS}$ is associated with a Jordan algebra, the following equality holds componentwise
\begin{equation}
C^{\tilde{I}\tilde{J}\tilde{K}}=C_{\tilde{I}\tilde{J}\tilde{K}}=const.\nonumber
\end{equation}

\section{Killing vectors of the hyper-isometry}

\setcounter{equation}{0}

The eight Killing vectors $k^X_\alpha$ that generate isometry group $SU(2,1)$ of the hyperscalar manifold are given by\cite{Ceresole:2001wi}
\begin{equation}\begin{array}{l}
\vec{k}_1=\left( \begin{array}{c}0\\1\\0\\0\end{array}\right),\quad
\vec{k}_2=\left( \begin{array}{c}0\\2\theta\\0\\1\end{array}\right),\quad
\vec{k}_3=\left( \begin{array}{c}0\\-2\tau\\1\\0\end{array}\right),\quad
\vec{k}_4=\left( \begin{array}{c}0\\0\\-\tau\\\theta\end{array}\right),\\
\vec{k}_5=\left( \begin{array}{c}V\\\sigma\\\theta/2\\\tau/2\end{array}\right),\quad
\vec{k}_6=\left( \begin{array}{c}2V\sigma\\\sigma^2-(V+\theta^2+\tau^2)^2\\\sigma\theta-\tau(V+\theta^2+\tau^2)\\\sigma\tau+\theta(V+\theta^2+\tau^2)\end{array}\right)\\
\vec{k}_7=\left( \begin{array}{c}-2V\theta\\-\sigma\theta+V\tau+\tau(\theta^2+\tau^2)\\\frac{1}{2}(V-\theta^2+3\tau^2)\\-2\theta\tau-\sigma/2\end{array}\right),\quad
\vec{k}_8=\left( \begin{array}{c}-2V\tau\\-\sigma\tau-V\theta-\theta(\theta^2+\tau^2)\\-2\theta\tau+\sigma/2\\\frac{1}{2}(V+3\theta^2-\tau^2)\end{array}\right).\label{tis}
\end{array}\end{equation}
The corresponding prepotentials are
\begin{equation}\begin{array}{l}
\vec{p}_1=\left( \begin{array}{c}0\\0\\-\frac{1}{4V}\end{array}\right),\quad
\vec{p}_2=\left( \begin{array}{c}-\frac{1}{\sqrt{V}}\\0\\-\frac{\theta}{V}\end{array}\right),\quad
\vec{p}_3=\left( \begin{array}{c}0\\\frac{1}{\sqrt{V}}\\\frac{\tau}{V}\end{array}\right),\quad
\vec{p}_4=\left( \begin{array}{c}-\frac{\theta}{\sqrt{V}}\\-\frac{\tau}{\sqrt{V}}\\\frac{1}{2}-\frac{\theta^2+\tau^2}{2V}\end{array}\right),\\
\vec{p}_5=\left( \begin{array}{c}-\frac{\tau}{2\sqrt{V}}\\\frac{\theta}{2\sqrt{V}}\\-\frac{\sigma}{4V}\end{array}\right),\quad
\vec{p}_6=\left( \begin{array}{c}-\frac{1}{\sqrt{V}}[\sigma\tau+\theta(-V+\theta^2+\tau^2)]\\\frac{1}{\sqrt{V}}[\sigma\theta-\tau(-V+\theta^2+\tau^2)]\\-\frac{V}{4}-\frac{1}{4V}[\sigma^2+(\theta^2+\tau^2)^2]+\frac{3}{2}(\theta^2+\tau^2)\end{array}\right),\\
\vec{p}_7=\left( \begin{array}{c}\frac{4\theta\tau+\sigma}{2\sqrt{V}}\\\frac{3\tau^2-\theta^2}{2\sqrt{V}}-\frac{\sqrt{V}}{2}\\-\frac{3}{2}\tau+\frac{1}{2V}[\sigma\theta+\tau(\theta^2+\tau^2)]\end{array}\right),\quad
\vec{p}_8=\left( \begin{array}{c}-\frac{3\theta^2-\tau^2}{2\sqrt{V}}+\frac{\sqrt{V}}{2}\\\frac{\sigma-4\theta\tau}{2\sqrt{V}}\\\frac{3}{2}\theta+\frac{1}{2V}[\sigma\tau-\theta(\theta^2+\tau^2)]\end{array}\right)
\end{array}\end{equation}

It is easier to see that the Killing vectors close to the $SU(2,1)$ algebra if they are recasted in the following combinations\footnote{This basis is chosen for convenience such that the generators $T_1, T_2, T_3$ and $T_8$ are the isotropy group of the point $(V, \sigma, \theta, \tau)=(1,0,0,0)$. The metric hyperscalar manifold becomes diagonal at this point. In all the theories that have hyper coupling, we will take this basis point $q^C$ for a possible candidate of the hyper-coordinates of a critical point.}

\begin{equation}
\begin{array}{rl}
SU(2)&\quad \left\{ \begin{array}{rcl}T_1&=&\frac{1}{4}(k_2-2k_8),\\T_2&=&\frac{1}{4}(k_3-2k_7),\\T_3&=&\frac{1}{4}(k_1+k_6-3k_4),\end{array}\right. \\[25pt]
U(1)&\quad \left\{ \begin{array}{rcl} T_8&=&\frac{\sqrt{3}}{4}(k_4+k_1+k_6),\end{array}\right.
\end{array}\qquad
\frac{SU(2,1)}{U(2)} \quad \left\{ \begin{array}{rcl}T_4&=&k_5,\\T_5&=&-\frac{1}{2}(k_1-k_6),\\T_6&=&-\frac{1}{4}(k_3+2k_7),\\T_7&=&-\frac{1}{4}(k_2+2k_8).\end{array}\right.
\label{Tvex}
\end{equation}
The Killing vectors $K_I^X$ are then given by $V_I^\alpha k^X_\alpha$ and the corresponding prepotentials $\vec{P}_I$ are $V_I^\alpha \vec{p}_\alpha$, where $V_I^\alpha$ are constants that determine which isometries are being gauged and what linear combination of vector fields being used. \footnote{In particular, 
\(
K_I^X=\left\{ \begin{array}{ll}
         T_1^X, T_2^X, T_3^X&\text{for } SU(2) \text{ gauging},\\
         V_I W^k T_k^X,\quad k=1,2,3,8& \text{for } U(1) \text{ gauging},\\
         V_I W^k T_k^X,\quad k=4,5,6,7& \text{for } SO(1,1) \text{ gauging},
        \end{array}
\right.
\)
where $V_I$ and $W^k$ are constants depending on the model.}

\end{document}